\documentclass{article}
\usepackage[english]{babel}
\usepackage[letterpaper,top=2cm,bottom=2cm,left=3cm,right=3cm,marginparwidth=1.75cm]{geometry}

\usepackage{quantikz}
\usepackage{amsmath}
\usepackage{amssymb}
\usepackage{bm} 
\usepackage{cite} 
\usepackage{graphicx}
\usepackage[colorlinks=true, allcolors=blue]{hyperref}
\usepackage{multirow} 
\usepackage{array} 
\usepackage{booktabs} 
\usepackage{subcaption} 
\usepackage[title]{appendix} 
\usepackage{authblk} 
\usepackage{amsfonts} 

\usepackage[normalem]{ulem} 

\AtBeginDocument{

}

\title{Grokking and epoch-wise double descent \\ in quantum neural networks}

\author[1,2]{Daniel Pranji\'{c}}
\author{Marco Roth}
\author{Christian Tutschku}

\affil[1]{Fraunhofer IAO, Nobelstraße 12, 70569 Stuttgart, Germany}
\affil[2]{Universität Stuttgart, Allmandring 35, 70569 Stuttgart, Germany}

\begin{document}

\maketitle

\begin{abstract} \noindent Grokking, the delayed transition from memorization to generalization, is a fundamental phenomenon in gradient-based learning, yet its dynamics within variational quantum machine learning (QML) remain largely unexamined. In this work, we report the empirical observation of both the grokking transition and epoch-wise double descent in a two-qubit quantum neural network (QNN) under a complete parameterization of the SU(4) manifold. We demonstrate that overparameterization via increased circuit depth improves the probability of successful generalization. Notably, these architectures frequently exhibit an epoch-wise double descent in test error, degrading at a critical epoch before recovering into a generalizing state. Crucially, we identify a \textit{generalization decay} in late-stage training, where the test error increases significantly despite a stagnant training loss. Bridging this behavior with algorithmic stability theory, our analysis reveals that this decay correlates with an unconstrained increase of the weight-norm, drifting away from sparse, phase-aligned harmonic solutions toward overfitted solutions in the Hilbert space. We analyze the underlying temporal dynamics of this transition, demonstrating how the onset of generalization is linked to optimization hyperparameters such as learning rate and weight decay. Finally, to mitigate late-stage decay, we introduce a weak explicit weight-norm regularization into the loss function. We demonstrate that this structural anchor stabilizes the post-grokking phase and permanently preserves generalization gains, providing a robust framework for training overparameterized quantum circuits.
\end{abstract}

\section{Introduction}

\noindent Understanding the phase transitions of overparameterized models remains a central challenge in contemporary machine learning. A hallmark of these dynamics is grokking: a sharp shift from memorization to generalization that occurs long after training performance has saturated~\cite{grokking_og}. While training and test losses often converge on similar time scales in standard regimes, grokking is characterized by a dramatic improvement in test accuracy thousands of epochs after the training loss has hit a plateau. Elucidating this phenomenon is of profound importance to general learning theory, as it fundamentally challenges classical generalization bounds by proving that meaningful representational shifts continue to evolve within completely saturated loss landscapes. Furthermore, decoding these delayed dynamics provides the necessary framework to understand why overparameterized architectures often lack the structural robustness to permanently maintain these generalization gains, leaving them highly vulnerable to severe performance decay during late-stage training.
Recent studies have analyzed this phenomenon through the lenses of representation learning~\cite{grokking_representation_learning}, weight-norm dynamics~\cite{grokking_weight_norm_dynamics}, singular learning theory~\cite{grokking_slt, grokking_slt2}, and algorithmic stability~\cite{stability_and_generalization, grokking_esann, grokking_do_we_really_need_new_theory}. More recently, investigations into tensor networks suggest that grokking may manifest as an entanglement transition, where generalization coincides with shifts in the representation's bond dimension and entropy~\cite{grokking_tensor_networks}. This intersection of quantum-inspired methods and machine learning hints at a deeper, quantum-mechanical dimension to these phase transitions, necessitating a systematic translation of these concepts from classical networks to true quantum architectures.

\noindent While significant attention in the quantum machine learning (QML) community has been devoted to the trainability of quantum neural networks (QNNs), specifically the Barren Plateau problem~\cite{qml_barren_plateaus}, the dynamics of these models \textit{after} the training loss has converged remain largely unexplored. In classical deep learning, it is well-understood that optimization continues along the flat valleys of the loss function, a process often associated with the discovery of more stable minima and simpler functions through implicit or explicit regularization. This high-capacity regime is closely linked to the double descent phenomenon~\cite{qml_double_descent, double_descent_general}, showing that generalization can paradoxically improve even as model complexity grows beyond the interpolation threshold. However, as QNNs approach the overparameterization limit~\cite{qnn_overparameterization}, the resulting models may lack the structural robustness necessary to maintain these generalization gains. To address this, the Lipschitz bound framework~\cite{qml_lipschitz} can be used to constrain the smoothness of the quantum model, providing a formal mechanism to regularize training dynamics within these high-dimensional and potentially degenerate landscapes.

\noindent In this work, we investigate the long-duration training dynamics of variational quantum circuits to demonstrate that QNNs fundamentally exhibit both grokking and epoch-wise double descent~\cite{double_descent_general, double_descent_epoch1, double_descent_epoch2, double_descent_epoch3}. To isolate these phenomena in a controlled setting, we employ a minimal 2-qubit QNN architecture that allows us to rigorously track landscape traversal without confounding multi-qubit overhead. We show that the resulting training plateau is not stationary but a dynamic regime where the model explores the degenerate manifold of the training loss, drifting between regions of varying algorithmic stability, Lipschitz bounds, weight-norms, and even entanglement. Finally, we demonstrate that the late-stage generalization decay is a direct consequence of the model's loss of structural robustness, and that applying a weight-norm regularization~\cite{qml_lipschitz} effectively anchors the QNN within a stable, generalizing regime. \\

\begin{figure}[t]
    \centering
    \begin{subfigure}{0.5\textwidth}
        \centering
        \includegraphics[width= \linewidth]{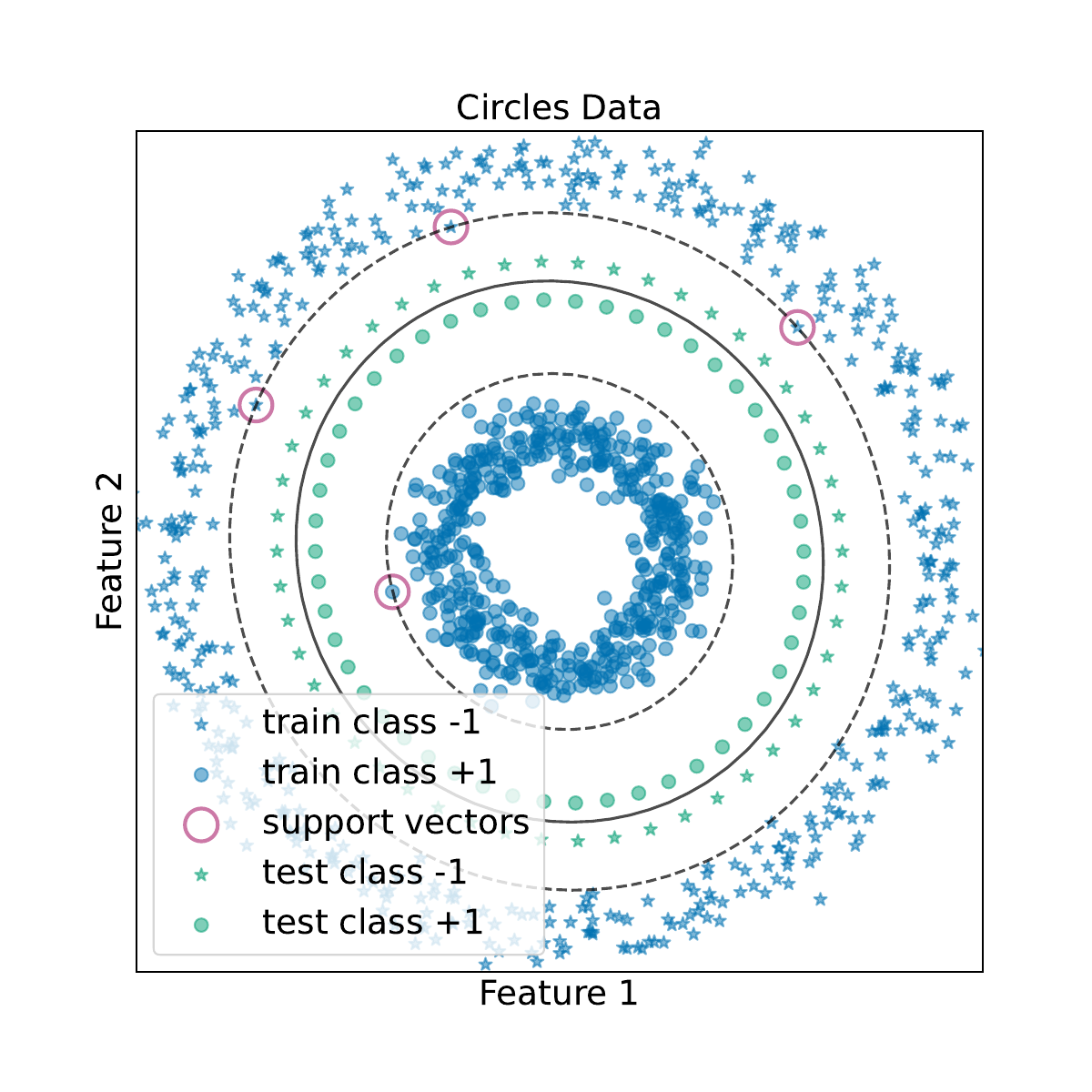}
        \caption{}
        \label{fig:data}
    \end{subfigure}%
    \begin{subfigure}{0.5\textwidth}
        \centering
        \includegraphics[width= \linewidth]{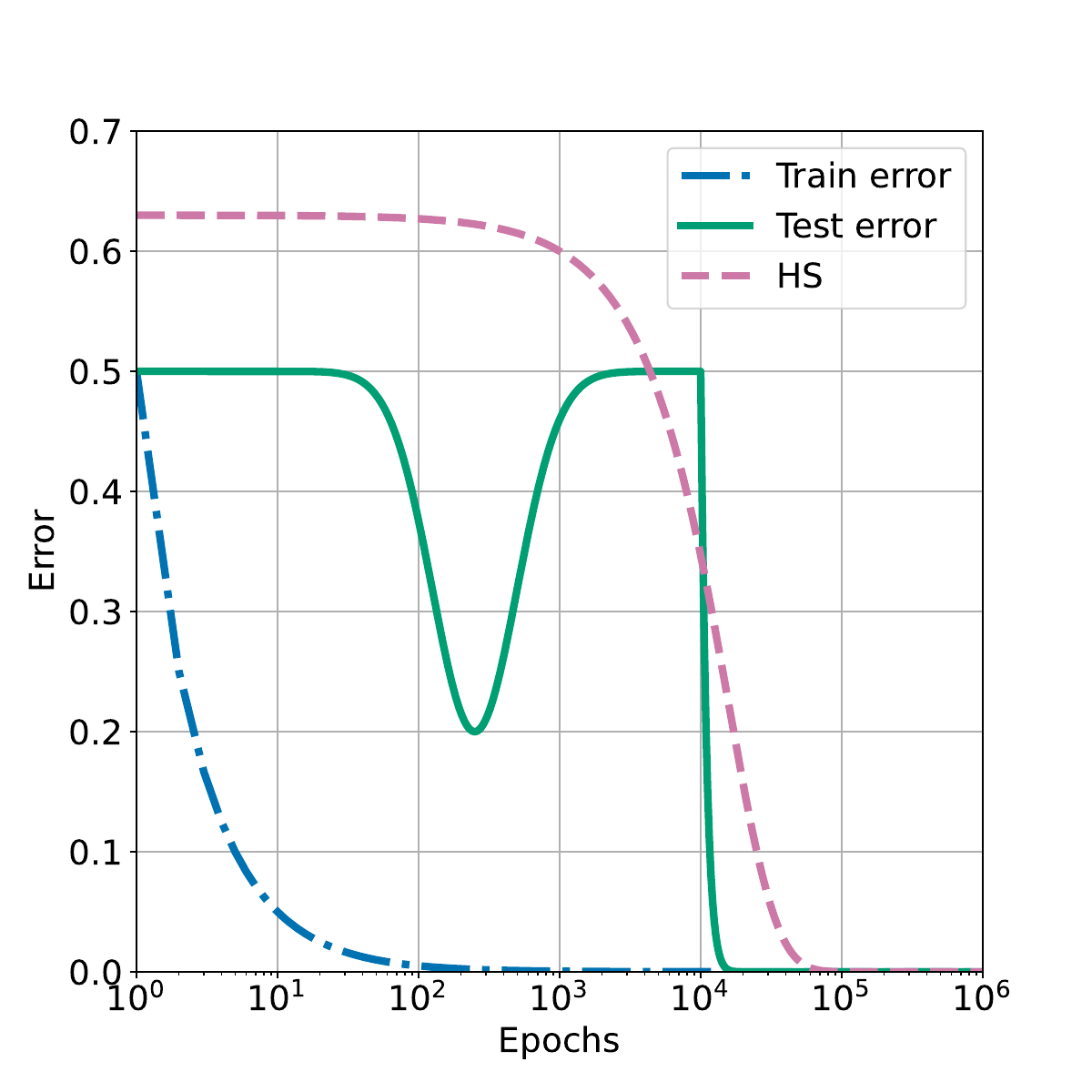}
        \caption{}
        \label{fig:sketch_grokking}
    \end{subfigure}

    \begin{subfigure}{\textwidth}
        \centering
        \includegraphics[width= \linewidth]{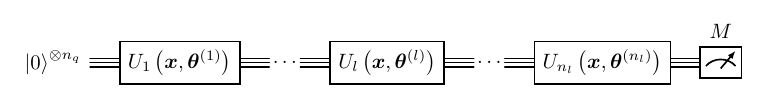}
        \caption{}
        \label{fig:sun_qnn}
    \end{subfigure}
    \caption{a) Diagram of the dataset described in~\autoref{sec:meth:dataset} showing random training points in two concentric circles (blue) and a ring of test points sitting closely to the SVM-defined decision boundary (green). The support vectors are highlighted in red circles. b) Sketch of a characteristic grokking loss curve with epoch-wise double descent. The model fits the train data quickly and enters the memorization phase. While the loss is effectively constant, the model weights can still change and influence other metrics, i.e. the hypothesis stability (HS). After thousands of epochs in the memorization phase, the model can show an abrupt transition, where the test error drops. This is called a grokking transition. In this case, grokking is accompanied by epoch-wise double descent, where the test error undergoes two non-monotonous transitions. c) Circuit diagram of the multi-layer QNN used throughout this work.}
\end{figure}

\noindent While the macroscopic signatures of grokking and double descent observed in this work share phenomenological similarities with classical deep learning, their manifestation in variational quantum circuits is fundamentally shaped by the non-Euclidean geometry of the parameter space. Optimization in a QNN does not take place in flat Euclidean space, but along the highly non-linear, compact parameters of a quantum gate manifold. This restriction to a compact and periodic parameter space forces a re-evaluation of standard optimization constraints. Crucially, our findings challenge the prevailing dogma surrounding the Barren Plateau phenomenon in QML, which traditionally mandates narrow parameter initializations ($\sigma_{\bm\theta} \to 0$) to preserve gradients. We demonstrate that in the overparameterized deep-circuit regime, broad initializations paired with strong weight decay do not trap the model indefinitely. Instead, they actively facilitate a highly structured representation search, compressing the timeline to generalization. This strengthens the position of overparameterization not as an optimization hazard, but as a vital resource for training QNNs. Rather than merely guaranteeing the minimization of empirical risk, the high-capacity regime provides the necessary parameter degrees of freedom for implicit regularization to systematically filter out non-generalizing solutions along the zero training loss manifold. \\

\noindent The primary contributions of this work are summarized as follows:

\begin{itemize}
    \item \textbf{Empirical observation of grokking in QNNs (\autoref{section:results:hero} and \autoref{section:results:groksistency}):} We provide empirical evidence of the grokking transition in SU(4) QNNs, identifying a distinct phase where the model shifts from memorization to generalization long after training convergence. We demonstrate the occurrence and stability of this grokking phase across different initializations and hyperparameter configurations. Both the ratio of runs that show the grokking transition and that are able to achieve low generalization error increase with the number of layers and trainable parameters while explicit weight-norm regularization can be used as an anchor against unwanted late-training stage parameter drifts.
    \item \textbf{Temporal dynamics of generalization (\autoref{sec:results:time}):} We quantify the "time until generalization" and generalization delay, revealing how learning rate and implicit regularization influence the onset of the generalization transition.
\end{itemize}

\section{Methodology}

In this section, we outline our experimental setup which has the goal of finding a minimal setup in which grokking and epoch-wise double descent can be studied in the context of QML. In~\autoref{sec:meth:dataset}, we introduce the dataset and demonstrate how generalization can be probed in a controlled setting of varying difficulty. In~\autoref{sec:meth:op_qnns}, we detail the overparameterized QNN architectures utilized throughout our experiments. In~\autoref{sec:meth:stability4qnns}, we introduce algorithmic stability theory for (Q)NNs as the theoretical framework to analyze these learning phases. A conceptual overview of the characteristic training trajectories can be found in~\autoref{fig:sketch_grokking}.

\subsection{Dataset and margin-based testing}
\label{sec:meth:dataset}

To rigorously evaluate the quality of the learned decision boundary, we follow the strategy from Reference~\cite{grokking_esann} for the training and test sets. The authors consider a binary classification task on a dataset of two-dimensional points. Their training dataset consists of linearly separable points. A hard-margin support vector machine (SVM) is applied to this data to identify the maximum margin solution. Then, the test data is placed close to the boundary. However, as established in foundational benchmarking studies~\cite{qml_benchmark, qml_benchmark2, qml_benchmark3}, standard quantum learners are fundamentally ill-suited for simple linear decision boundaries due to the inherent non-linearity of mapping data into the high-dimensional Hilbert space. Consequently, we construct a non-linear classification task in the following way

\begin{itemize}
    \item Training Set $\mathcal{D}_\mathrm{train} = \mathcal{X} \times \mathcal{Y} = \left\{ \left(\bm{x}^{(i)}, y^{(i)}\right) \right\}_{i=1}^N$: The points $\bm{x}^{(i)} \in \mathcal{X}$ are sampled randomly from two concentric circular classes with labels $y^{(i)} \in \mathcal{Y}$, see~\autoref{fig:data}.
    \item Test Set $\mathcal{D}_\mathrm{test}$: To specifically probe the model's structural robustness, the test points are distributed evenly with a small distance $\varepsilon > 0$ along the optimal decision boundary. This boundary is pre-calculated using a classical hard-margin SVM with a quadratic kernel~\cite{svm_steinwart}.
\end{itemize}

\noindent Because the hard-margin SVM boundary is defined by the maximum-margin criterion, it represents a unique and optimal boundary for separating the concentric classes. By placing the test data close to the SVM boundary, we create a highly sensitive diagnostic for the QNN. In this regime, even a small change of the weights that maintains zero training loss can shift the decision boundary enough to misclassify the margin points.

\subsection{Overparameterized QNNs}
\label{sec:meth:op_qnns}

We embed the two-dimensional data samples into a 15-dimensional feature vector

\begin{equation}
    \phi(\bm{x}) = (x_1, x_2, 1, 1, \dots, 1) \in\mathbb{R}^{15} \, .\label{eq:feature_map}
\end{equation}

\noindent This embedding represents a deliberate architectural choice. Despite utilizing a two-dimensional input dataset, it allows us to parameterize the complete SU(4) group within a single layer, thereby capturing the full expressive capacity of the two-qubit Hilbert space. While more complex, non-linear quantum feature maps exist, this linear padding is deliberately chosen to isolate the training dynamics on the SU(4) manifold without introducing data-wrapping non-linearities. The SU(4) unitary is parameterized by multiplying these \textit{padded} features to the 15 generators $\{ G_k \}_{k=1}^{15}$ of the $\mathfrak{su}(4)$ Lie algebra

\begin{equation}
    U^{(l)}(\bm{x}, \bm{\theta}) = \exp\left( i \sum\limits_{k=1}^{15} \, \theta^{(l)}_k \, \phi_k(\bm{x}) \, G_k \right) \, , \label{eq:su4_unitary}
\end{equation}
where $\theta^{(l)}_1, \dots, \theta^{(l)}_{15}$ are the trainable weights of the $l$-th layer sampled from a Gaussian $\mathcal{N}(0, \sigma_{\bm\theta})$ centered at zero with standard deviation $\sigma_{\bm\theta}$. The model output is the expectation value of a measurement operator $M$

\begin{equation}
    f(\bm{x}; \bm\theta) = \bra{\psi(\bm{x}, \bm\theta)}M\ket{\psi(\bm{x}, \bm\theta)} \, , \label{eq:model}
\end{equation}
where $\ket{\psi(\bm{x},\bm\theta)} = \prod_{l=1}^{n_l} U^{(l)}\left(\bm{x}, \bm\theta^{(l)}\right) \ket{00}$ and $n_l$ denotes the total number of layers of the model, see the corresponding circuit diagram in~\autoref{fig:sun_qnn}. The optimization objective is to minimize the following regularized loss

\begin{equation}
    \mathcal{L}(\bm\theta) = \frac{1}{N} \sum\limits_{(\bm{x},y) \in \mathcal{D}_\mathrm{train}} \, \ell\left( y, f(\bm{x}; \bm\theta) \right) + \lambda \, \sum\limits_{k=1}^{15} \, \sum\limits_{l=1}^{n_l} \left( \theta^{(l)}_k \right)^2  \, , \label{eq:objective}
\end{equation}
where $\ell(y, \hat{y}) = (y-\hat{y})^2$ is the loss between the true ($y$) and predicted label ($\hat{y} = f(\bm{x};\bm\theta)$), and $\lambda \ge 0$ is the weight-norm regularization strength. Furthermore, we will track the Lipschitz bound of $f(\bm{x}; \bm\theta)$ given by~\cite{qml_lipschitz} 

\begin{equation}
    L(\bm\theta) = 2 \sum\limits_{k=1}^{2} \, \sum\limits_{l=1}^{n_l} \left| \theta^{(l)}_k \right| \, , \label{eq:lipschitz_bound}
\end{equation}
in our implementation with measurements in the $z$-basis ($M = Z \otimes Z$, where $Z$ is the Pauli-Z matrix and the $G_k$'s are just Pauli strings). The indices $k=1,2$ belong to the two non-padded, data-carrying dimensions of the feature vector in~\autoref{eq:feature_map}.

\noindent In our binary classification setting, we obtain the classifier by taking the sign of $f$. Note, that the Lipschitz bounds is a bound of the generalization error of $f(\bm{x}; \bm\theta)$ (see more detailed version with proof in Ref.~\cite{qml_lipschitz})

\begin{equation}
    \left|\mathcal{R}\left[f(\bm\theta)\right] - \mathcal{R}_\mathrm{emp}\left[f(\bm\theta)\right] \right| \le C_1 L(\bm\theta) + \frac{C_2}{\sqrt{N}} \, ,
    \label{eq:lipschitz_generalization_error_bound}
\end{equation}
for some constants $C_1, C_2 > 0$ with expected risk $\mathcal{R}\left[f(\bm\theta)\right]= \int_\mathcal{\mathcal{X} \times \mathcal{Y}} \, \ell(y, f(\bm{x};\bm\theta)) \, \mathrm{d}p(\bm{x},y)$ for an unknown probability distribution $p(\bm{x},y)$ and empirical risk $\mathcal{R}_\mathrm{emp}\left[f(\bm{x};\bm\theta)\right]=\sum_{k=1}^{N} \ell(y_k, f(x_k; \bm\theta))/N$. From~\autoref{eq:lipschitz_bound} it becomes apparent that the Lipschitz bound depends exclusively on the weights coupled to the data-dependent generators within the unitaries defined in~\autoref{eq:su4_unitary}. In our specific architecture, this implies that only two parameters per layer actively contribute to the evaluation of the Lipschitz bound. Note that after reaching the zero-training-error manifold, the model can continue to accumulate weight-norm during extended training, leading to a subsequent decay in generalization. Because this weight accumulation in late-stage training affects all the weights, we have chosen a global regularization scheme across all weights in~\autoref{eq:objective}. Alternatively, one could penalize the Lipschitz-active and remaining weight-norms independently by decomposing the regularization term in~\autoref{eq:objective} using two distinct hyperparameters, $\lambda_{L}, \lambda_{\neg L} \ge 0$. \\

\noindent To minimize the regularized loss objective in~\autoref{eq:objective}, we employ the AdamW optimizer~\cite{adamw}, a variant of the Adam algorithm that decouples weight decay from the gradient update. Given the vastness of the reachable manifold ($15 n_l$ parameters versus two input dimensions), AdamW is chosen for its ability to navigate flat loss landscapes while suppressing the weight-norm. This drives the model towards lower-weight and smoother solutions and serves as a critical mechanism for inducing the grokking transition in the overparameterized regime.

\subsection{Algorithmic stability theory for (Q)NNs}
\label{sec:meth:stability4qnns}

\noindent To formally characterize the generalization dynamics during the grokking transition, we utilize the framework of algorithmic stability theory~\cite{grokking_esann, stability_and_generalization}. Algorithmic stability provides a distribution-agnostic bound on the generalization gap during training. More concretely, the expected generalization gap is upper-bounded by the \textit{uniform hypothesis stability} (HS) $\beta$, such that the expected risk $\mathcal{R}$ satisfies
\begin{equation}
    \mathbb{E}[\mathcal{R}] \le \mathbb{E}[\mathcal{R}_\mathrm{emp}] + \beta \, , \label{eq:def_beta}
\end{equation}
where $\mathcal{R}_\mathrm{emp}$ denotes the empirical risk and the expectation $\mathbb{E}$ is evaluated with respect to the training dataset $\mathcal{D} \in \mathcal{Z}^N$ sampled from the data domain $\mathcal{Z} = \mathcal{X} \times \mathcal{Y}$. To formally define $\beta$, let $A_\mathcal{H}: \mathcal{Z}^N \to \mathcal{F}$ represent a learning algorithm governed by hyperparameters $\mathcal{H}$ that maps an arbitrary dataset $\mathcal{D} \in \mathcal{Z}^N$ to a predictor function $f: \mathcal{X} \to \mathcal{Y}$ within a hypothesis family $\mathcal{F}$. Then, the algorithm $A_\mathcal{H}$ exhibits \textit{uniform} HS $\beta$ with respect to a loss function $\ell$ if the absolute change in loss incurred by omitting any single training sample $z_i \in \mathcal{D}$ is bounded across all possible training datasets $\mathcal{D}$, all arbitrary evaluation points $z \in \mathcal{Z}$ and all sample indices $i \in \{ 1, \dots, N \}$~\cite{stability_and_generalization}
\begin{equation}
    \sup_{\mathcal{D} \in \mathcal{Z}^N, z \in \mathcal{Z}, i\in \{ 1, \dots, N \}} \left| \ell\big(A_{\mathcal{H}}(\mathcal{D}), z\big) - \ell\big(A_\mathcal{H}(\mathcal{D} \setminus \{z_i\}), z\big) \right| \le \beta \, .
    \label{eq:uniform_stability}
\end{equation}
While uniform HS allows for the derivation of rigorous exponential generalization bounds via concentration inequalities~\cite{stability_and_generalization}, evaluating a pointwise supremum over an entire non-linear parameter landscape during runtime is analytically and computationally intractable. Therefore, we track the expected \textit{leave-one-out} HS over the data distribution, defined as 
\begin{equation}
    \beta_\mathrm{loo} = \mathbb{E}_{\mathcal{D} \in \mathcal{Z}^N, z \in \mathcal{Z}} \, \left| \ell(A_\mathcal{H}(\mathcal{D}), z) - \ell( A_\mathcal{H}(\mathcal{D} \setminus \{z_i\}), z ) \right| \, . \label{eq:hypothesis_stability}
\end{equation}
Since the strict uniform stability requirement acts as an upper bound on all average-case variations ($\beta_{\text{loo}} \le \beta$), tracking empirical metrics proportional to this expected functional deviation serves as a mathematically sound proxy for monitoring the (Q)NNs HS. \\
In our subsequent empirical evaluations (cf. \autoref{section:results:hero}), we explicitly compute and track $\beta_\mathrm{loo}$ from~\autoref{eq:hypothesis_stability} directly. \footnote{While tracking this exact loss-based metric provides an unadulterated view of the model's algorithmic stability during runtime, performing exhaustive leave-one-out retraining is computationally prohibitive and scales linearly with the dataset size $N$. 
To point toward more scalable diagnostics, we note that this expected functional variance can alternatively be approximated analytically without explicit dataset retraining by analyzing the spectrum of the (Q)NN's representation space~\cite{grokking_esann}
\begin{equation}
    \begin{aligned}
    \beta_{\text{loo}} &\propto \text{cond}(G_f) \, , \\
    G_f &= \Phi_f^T \Phi_f \, , \\
    \Phi_f(\bm{\theta}) &= \big(f(\bm{x}^{(1)}; \bm{\theta}), \dots, f(\bm{x}^{(N)}; \bm{\theta})\big)^T \, ,     
    \end{aligned}
    \label{eq:gramian_prop} 
\end{equation}
where $\text{cond}(G_f)$ denotes the condition number of the representation Gramian matrix $G_f$, constructed from the (Q)NN prediction vectors $\Phi_f(\bm{\theta})$.}

\section{Results}

\autoref{section:results:hero} details a concrete example of the grokking transition and its accompanying epoch-wise double descent in a single training run. \autoref{section:results:groksistency} analyzes the grokking behavior statistically across many independent runs to evaluate consistency and hyperparameter sensitivity. \autoref{sec:results:time} characterizes the temporal scaling properties of the generalization transitions.

\subsection{Empirical observation of the grokking transition and epoch-wise double descent}
\label{section:results:hero}

\begin{figure}[h]
    \centering
    \begin{subfigure}{0.47\textwidth}
        \includegraphics[width=\linewidth]{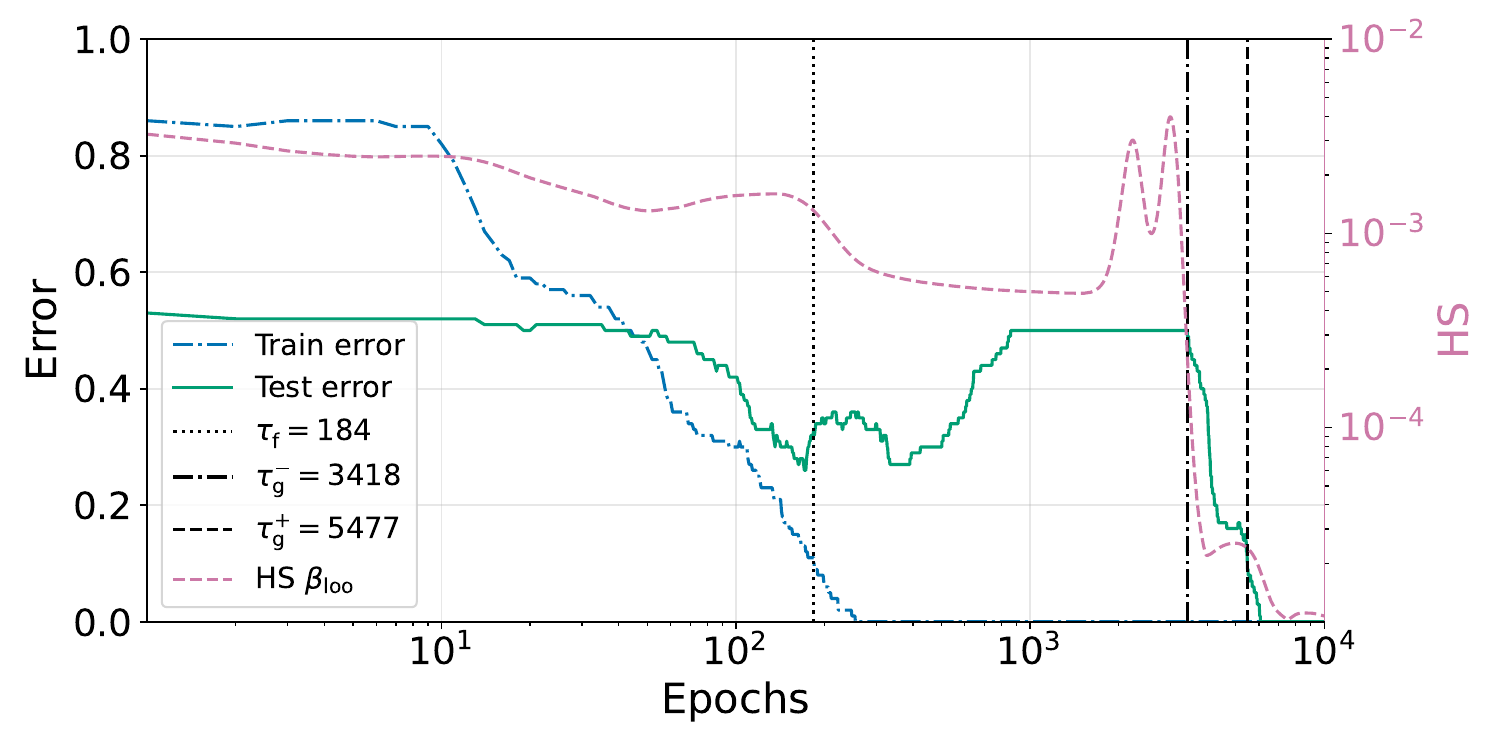}
        \caption{}  
        \label{fig:hero_history}
    \end{subfigure}%
    \begin{subfigure}{0.52\textwidth}
        \includegraphics[width=\linewidth]{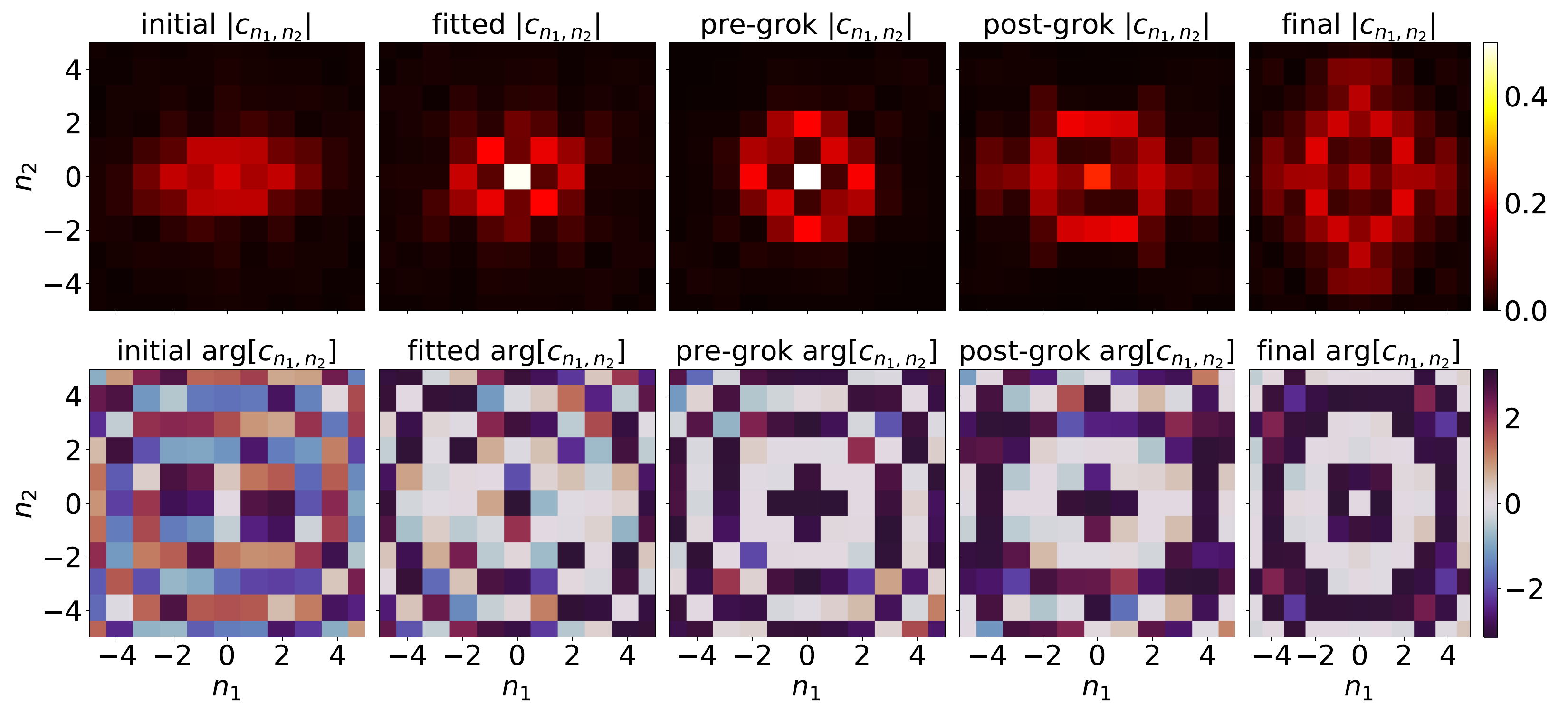}
        \caption{}
        \label{fig:hero_fourier}
\end{subfigure}
    \caption{Empirical observation of grokking and post-transition stability. The training history for the model in~\autoref{eq:model} trained on the dataset defined in~\autoref{sec:meth:dataset} is shown in (a). Notably, the hypothesis stability $\beta_\mathrm{loo}$ undergoes a sharp drop around the grokking transition from epoch 3000 to $\tau_\mathrm{g}^+ = 5477$. Parameters $n_l=3, \, | \mathcal{D}_\mathrm{test}|=|\mathcal{D}_\mathrm{train}|=100, \, \varepsilon=0.1, $ learning rate $\eta = 10^{-3}$, weight decay $\lambda_W = 10^{-5}, \sigma_{\bm\theta} = 1$. (b) Heatmap of Fourier coefficients $c_{n_1, n_2}(\theta_\tau)$ during different stages of training, $\tau \in \left\{ 0, \tau^{ }_\mathrm{f}, \tau^{\pm}_{\mathrm{g}}, 10^5 \right\}$. $\tau^{-}_\mathrm{g}$ is the last epoch of the memorization phase and $\tau^{+}_\mathrm{g}$ is the first epoch where the test error drops below 0.1.}
    \label{fig:hero}
\end{figure}

\noindent We report both grokking and epoch-wise double descent within the model framework defined in~\autoref{eq:model}. As shown in~\autoref{fig:hero_history}, the QNN's learning trajectory undergoes three distinct training-stages: initial fitting, a prolonged memorization plateau, and a sudden generalization transition (grokking). Initially, the model rapidly minimizes training error, achieving train error below 0.1 at the fitting epoch $\tau_\mathrm{f} = 184$. This first descent reduces the test error to a local minimum of approximately $0.3$, failing to achieve meaningful generalization. Following this, from $\tau_\mathrm{f}$ to the memorization exit epoch $\tau_\mathrm{g}^- = 3418$ (marking the final epoch of the static memorization phase), the model enters a stable memorization phase. Throughout this plateau, both training and test errors remain entirely static. From an exterior perspective, the model's predictive behavior appears frozen. However, evaluating the internal parameter dynamics reveals that the QNN is undergoing continuous, structured reorganization during this phase. This is captured by the hypothesis stability $\beta_\mathrm{loo}$ (see~\eqref{eq:def_beta}), which evolves during the memorization plateau before undergoing a sharp drop precisely coinciding with the grokking transition between $\tau_\mathrm{g}^-$ and the generalization epoch $\tau_\mathrm{g}^+ = 5477$, defined as the first training epoch where the test error drops below $0.1$. At this critical juncture, the second descent occurs, and the test error abruptly plummets to zero. \\

\noindent To uncover the mechanism driving this transition,~\autoref{fig:hero_fourier} tracks the evolution of the Fourier coefficients $c_{n_1, n_2}(\theta_\tau)$ of the QNN output function $f(\bm{x};\theta_{\tau})$ on the two-dimensional input domain across training epochs. At initialization ($\tau=0$), the weight magnitudes are diffuse and unstructured, with an isotropic distribution centered at the origin, accompanied by a chaotic, disordered phase argument. When the training data is first fitted ($\tau_\mathrm{f}$), the Fourier magnitude $|c_{n_1, n_2}|$ concentrates heavily on a centralized cluster, indicating that low-frequency, localized features are leveraged to memorize the training samples. Crucially, during the seemingly static memorization phase ($\tau^{ }_\mathrm{f} \to \tau_\mathrm{g}^-$), the Fourier coefficients begin to rearrange. The pre-grok heatmap ($\tau_\mathrm{g}^-$) reveals an evacuation of the central density and the initial formation of an outer ring structure. This indicates that even though the loss is stationary, weight decay and gradient dynamics are systematically filtering out non-generalizing modes. During the grokking transition ($\tau_\mathrm{g}^+$), this sparse geometric structure solidifies abruptly. The magnitude heatmap manifests a clean, symmetric ring-like topology, while the phase argument resolves into highly ordered, concentric bands. This implies that generalization is fundamentally characterized by a transition from a dense, unaligned representation to a sparse, phase-aligned harmonic solution. Beyond the transition ($\tau = 10^5$), this sparse Fourier representation is preserved and further crystallized, verifying the post-transition stability of the generalized solution. \\

\noindent This structural re-alignment can be understood by examining the \textit{idealized} target function of the dataset. Given that the optimal decision boundary is a circle of radius $R$ separating labels $y \in \{-1, +1\}$ inside a square domain of side length $L$ centered at the origin, the ideal classifier maps to a radially symmetric step function $f(\mathbf{x}) = 2 \cdot \mathbb{I}(r \le R) - 1$, where $r = \lVert\mathbf{x}\rVert$ denotes the radial distance and $\mathbb{I}$ is the indicator function. As derived in~\autoref{app:fourier_computation} using a high-density grid approximation of $N$ uniform samples, evaluating the discrete Fourier transform of this target function yields an analytical baseline for the discrete coefficients
\begin{equation}\label{eq:bessel_discrete_main}
c_{n_1, n_2} \approx \begin{cases} 
2 \left( \dfrac{\pi R^2}{L^2} \right) - 1 & \text{if } n_1 = 0 \text{ and } n_2 = 0, \\[12pt]
\dfrac{2R}{L} \dfrac{J_1\left(2\pi \dfrac{R}{L} \sqrt{n_1^2 + n_2^2}\right)}{\sqrt{n_1^2 + n_2^2}} & \text{otherwise},
\end{cases}
\end{equation}
where $n_1, n_2 \in \mathbb{Z}$ are the discrete harmonic frequency indices, and $J_1$ is the first-order Bessel function of the first kind. Aside from a constant vertical scale shift localized entirely at the zero-frequency origin ($n_1=0, n_2=0$), the frequency domain landscape is mathematically dominated by a classic Airy-like diffraction pattern~\cite{airy_og, airy_book}. This pattern is characterized by concentric rings of oscillating intensity that decay asymptotically as $\mathcal{O}(\lVert\mathbf{n}\rVert^{-3/2})$. This analytical baseline explains both the emergence of the well-defined rings and the fading outward amplitude observed in the post-grokking magnitude spectrum in~\autoref{fig:hero_fourier}.
Crucially, this mathematical structure also provides a rigorous explanation for the complex phase dynamics ($\arg(c_{n_1, n_2})$) across the training. Because the ideal target function is purely real-valued and perfectly centered, its theoretical Fourier coefficients are strictly real numbers. In the complex plane, a purely real value is forced into a binary phase profile, exhibiting a phase of exactly $0$ where the expression is positive, and flipping sharply to $\pm\pi$ where it is negative. The concentric phase bands visible in the post-grokking regime are therefore a direct visual signature of the quantum network successfully collapsing its chaotic, early-stage complex parameter overlays into a low-frequency manifold aligned with the alternating real roots of the classical Bessel function.

\subsection{Occurrence and stability of the grokking phase}
\label{section:results:groksistency}

\begin{figure}[h]
    \centering
    \begin{subfigure}{0.33\textwidth}
        \includegraphics[width=\linewidth]{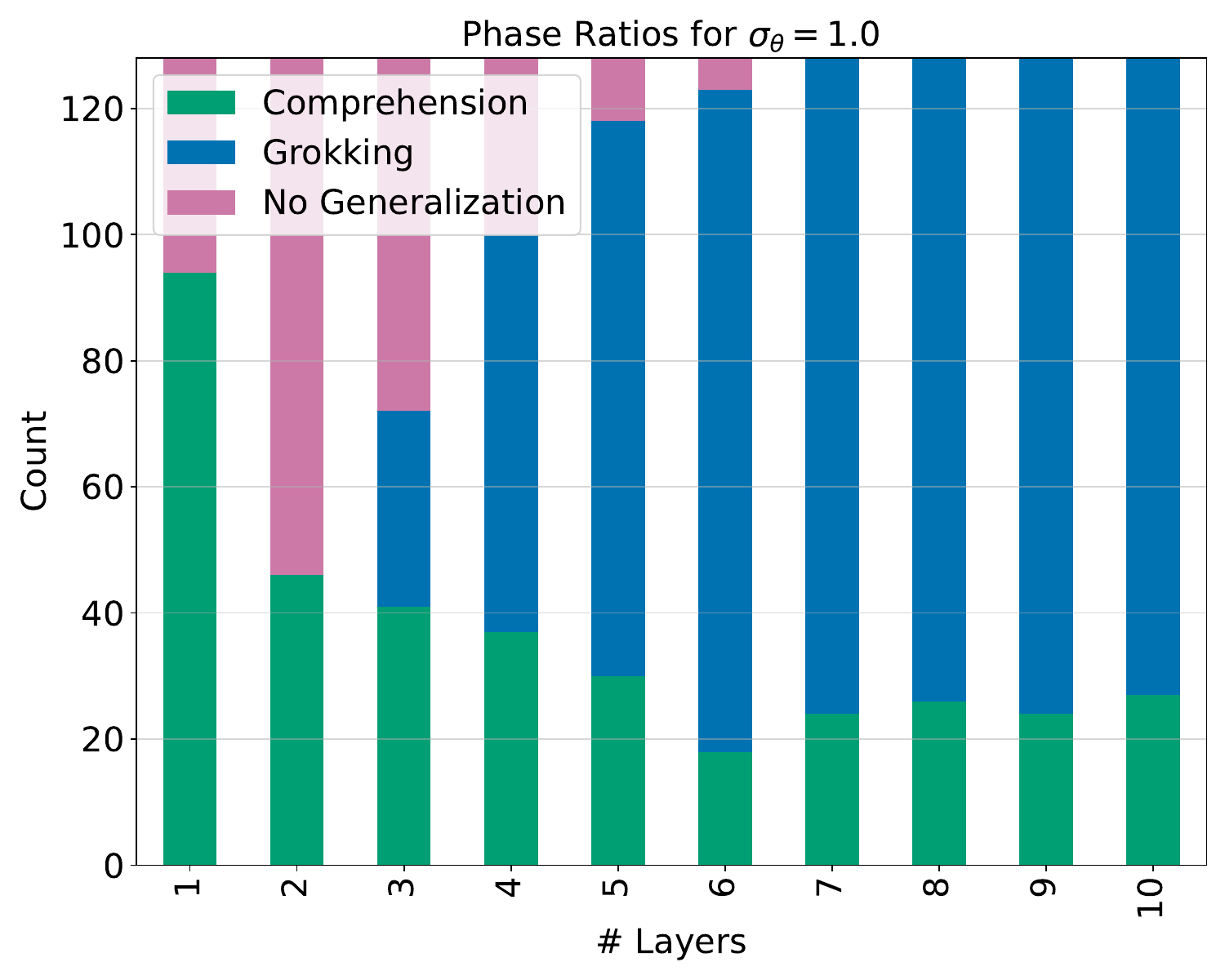}
        \caption{}
        \label{fig:groksistency_phase_ratios}
    \end{subfigure}%
    \begin{subfigure}{0.33\textwidth}
        \includegraphics[width=\linewidth]{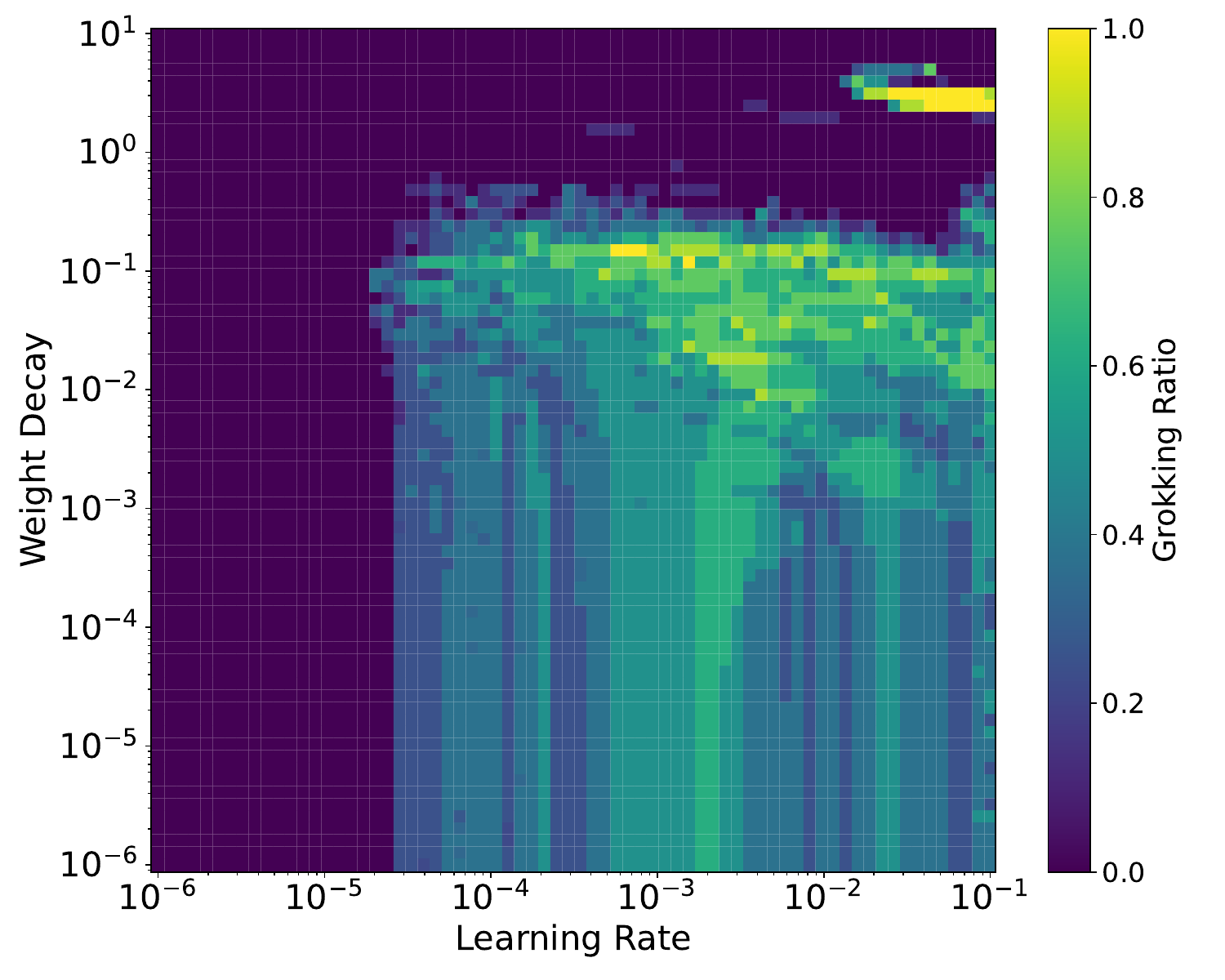}
        \caption{}
        \label{fig:groksistency_lr_wd_grok}
    \end{subfigure}%
    \begin{subfigure}{0.33\textwidth}
        \includegraphics[width=\linewidth]{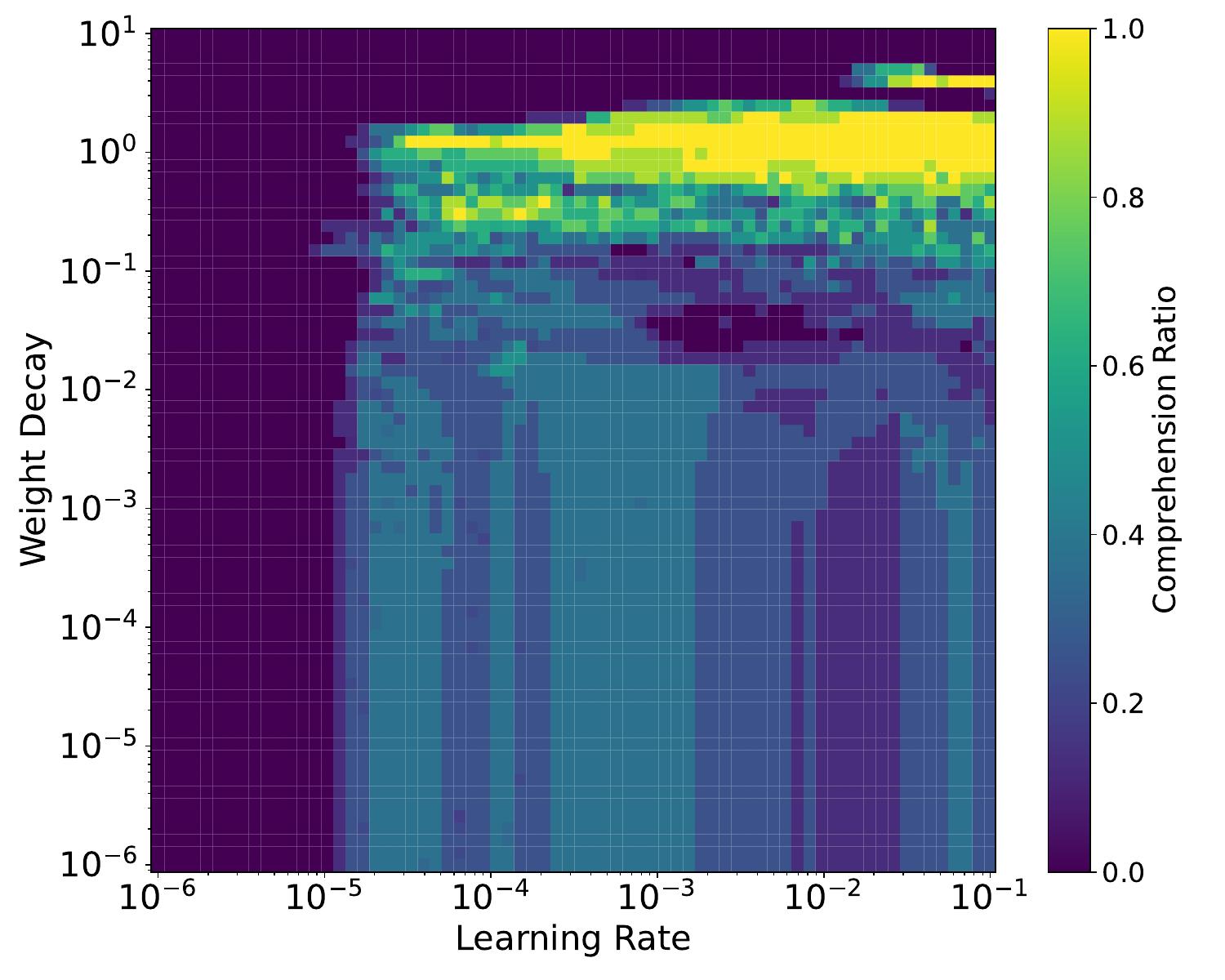}
        \caption{}
        \label{fig:groksistency_lr_wd_comp}
    \end{subfigure}
    \caption{a) Barplot of the 128 independent training runs of the quantum learner for each number of layers $n_l \in \{ 1, \dots 10 \}$. The weights $\bm\theta$ are initialized according to a Gaussian with spread $\sigma_{\theta} = 1.0$ and $|\mathcal{D}_{\mathrm{train}}|=100=|\mathcal{D}_\mathrm{test}|, \, \varepsilon=0.1, \, \lambda=0$ and $10^5$ epochs. For higher $n_l$ the ratio of runs with a successful generalization increases. For $n_l \ge 7$ no unsuccessful runs have been observed. b) Heatmap of the grokking ratio and c) heatmap of the comprehension ratio, both aggregated over 8 independent training runs of the quantum learner for $n_l = 3$ and $\sigma_{\bm\theta}=2\pi$, evaluated accross learning rates $\eta \in \left[10^{-6}, 10^{-1} \right]$ and weight decay $\lambda_W \in \left[10^{-6}, 10\right]$.}
    \label{fig:groksistency_phases}
\end{figure}

\begin{table}
\centering
\caption{Categorization of learning phases. The time gap condition for distinguishing grokking from comprehension chosen in this work is $\tau_\mathrm{g} / \tau_\mathrm{f} > 10$. The time to generalization has to be much larger than the time to fit the data $\tau_\mathrm{g} \gg \tau_\mathrm{f}$. In other resources the time gap condition is typically chosen as an absolute difference $\tau_\mathrm{g}-\tau_\mathrm{f} = 10^3$~\cite{grokking_representation_learning}.}
\begin{tabular}{lcccc}
\toprule
\textbf{Phase} & \textbf{Train error $< 10\%$} & \textbf{Test error $< 10\%$} & \textbf{Time Gap\textsuperscript{*}} \\ \midrule
Comprehension  & Yes & Yes & Small ($< 10$) \\
Grokking       & Yes & Yes & Large ($> 10$) \\
Memorization   & Yes & No  & N/A \\
Confusion      & No  & No  & N/A \\ \bottomrule
\multicolumn{4}{l}{\footnotesize \textsuperscript{*}Time Gap: $\underbrace{\tau(\text{test error} < 10\%)}_{\tau_\mathrm{g}} / \underbrace{\tau(\text{train error} < 10\%)}_{\tau_\mathrm{f}}$}
\end{tabular}
\label{tab:phase_categorization}
\end{table}

\noindent To systematically investigate the occurrence and stability of the grokking transition, we conduct an extensive multi-seed analysis. We initialize the model across 128 independent random seeds (sampled from a Gaussian distribution with zero mean and standard deviation $\sigma_{\bm\theta} = 1.0$) while varying the model depth across $n_l \in \{1, \dots, 10\}$. Each resulting trajectory is categorized according to the programmatic criteria detailed in~\autoref{tab:phase_categorization}. For empirical aggregation in our figures, the individual failures to generalize (comprising both the localized \textit{Memorization} and \textit{Confusion} phases) are consolidated under the label \textit{No Generalization}. \\

\noindent Our statistical results, illustrated  in~\autoref{fig:groksistency_phase_ratios}, reveal a striking architectural dependence on QNN depth. For shallow instances ($n_l < 3$), grokking is completely absent within the allotted training horizon ($10^5$ epochs). Instead, these configurations are predominantly split between clean comprehension and generalization failure. However, as depth increases beyond $n_l = 2$, we observe a robust, positive correlation with both the occurrence rate of the grokking phase and the global generalization success rate. When the QNN depth reaches $n_l \ge 7$, the stochastically driven failure modes associated with random initialization are entirely suppressed. All of the 128 independent runs successfully converge to a generalizing solution. This suggests that overparameterization via layer depth fundamentally smoothens the optimization landscape, mitigating the risks of unfavorable initial seed trajectories~\cite{qnn_overparameterization, nn_overparameterization}. \\

\noindent Beyond architectural constraints, hyperparameters like the learning rate $\eta$ and weight decay $\lambda_\mathrm{W}$ exert a profound influence on optimization dynamics and the resulting learning phases. To map out these phases and demonstrate that grokking is not merely a statistical fluke, we execute a dense grid search over $(\eta, \lambda_W)$ utilizing a 3-layer model across 8 independent seeds per coordinate, tracking both the grokking ratio (\autoref{fig:groksistency_lr_wd_grok}) and comprehension ratio (\autoref{fig:groksistency_lr_wd_comp}). Note that we employ a larger initialization $\sigma_{\bm\theta}=2\pi$ here so that the regularizing effect of the weight decay $\lambda_W$ is more pronounced. Together, these phase diagrams highlight a highly structured, contiguous landscape where successful generalization is remarkably robust across wide parameter regions rather than being an isolated algorithmic artifact. Notably, \textit{Comprehension} dominates in areas of high weight decay ($\lambda_W \gtrsim 1$). Crucially, this \textit{Comprehension} zone directly borders a massive \textit{Grokking} domain that occupies the central and moderate learning rate space ($\eta \in [10^{-4}, 10^{-1}]$), remaining resilient across several orders of magnitude of weight decay ($\lambda_W \in [10^{-6}, 10^{-1}]$). A \textit{No Generalization} phase only emerges at the extreme upper (high $\lambda_W \to 10$) and leftmost boundaries ($\eta < 10^{-5}$), where either excessive weight contraction overpowers the delicate representation alignment required for learning or the epoch budget is insufficient, leaving the model stranded in non-generalizing states. \\
While these parameters govern the onset of generalization, long-duration training reveals a severe vulnerability. Specifically, unregularized overparameterized models frequently exhibit post-transition drift, causing the learned representation to degrade back into high-complexity, non-generalizing solutions. As statistically detailed via the violin distributions in~\autoref{app:groksistency_lambda} (cf.~\autoref{fig:groksistency_l2}), introducing a weak, explicit weight-norm regularization ($\lambda=10^{-4}$) serves as a critical structural anchor. This penalty does not alter the absolute speed or delayed nature of the grokking transition itself ($\tau_\mathrm{g}/\tau_\mathrm{f}$). Instead, it actively suppresses late-stage generalization decay and permanently locks the QNN into the sparse, generalized state once discovered.

\subsection{Temporal characterization of the grokking transition}
\label{sec:results:time}

\begin{figure}[h]
    \centering
    \begin{subfigure}{0.33\textwidth}
        \includegraphics[width=\linewidth]{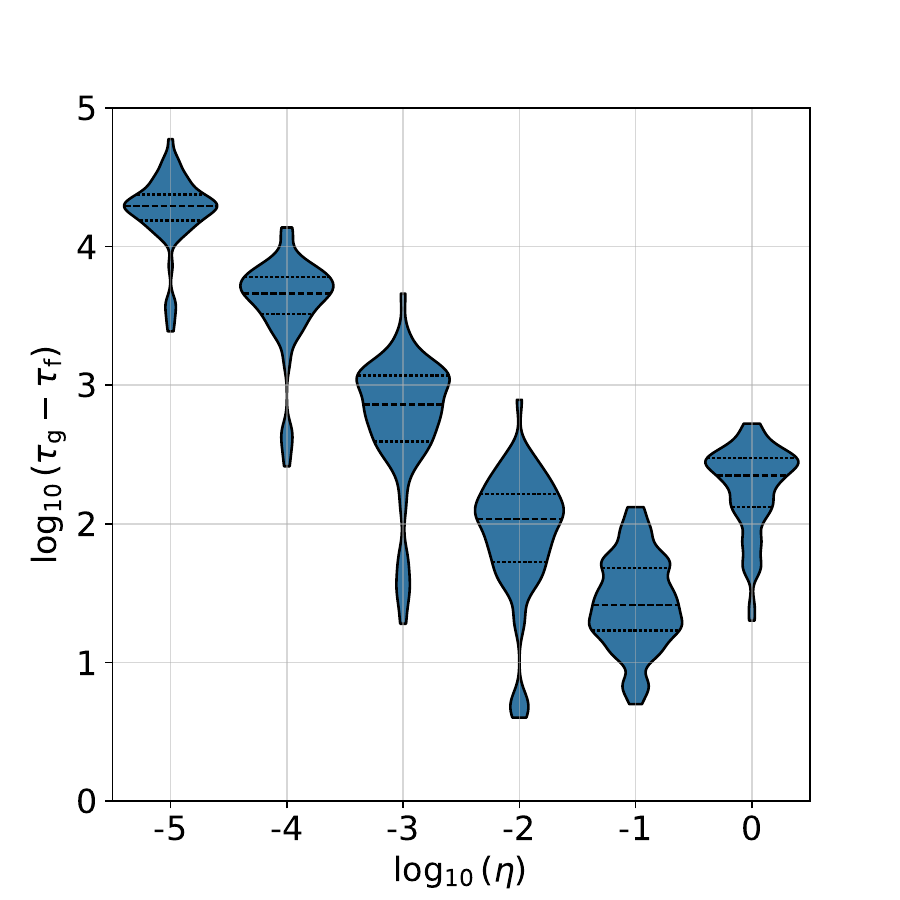}
        \caption{}
        \label{fig:temporal_dynamics_lr}
    \end{subfigure}%
    \begin{subfigure}{0.33\textwidth}
        \includegraphics[width=\linewidth]{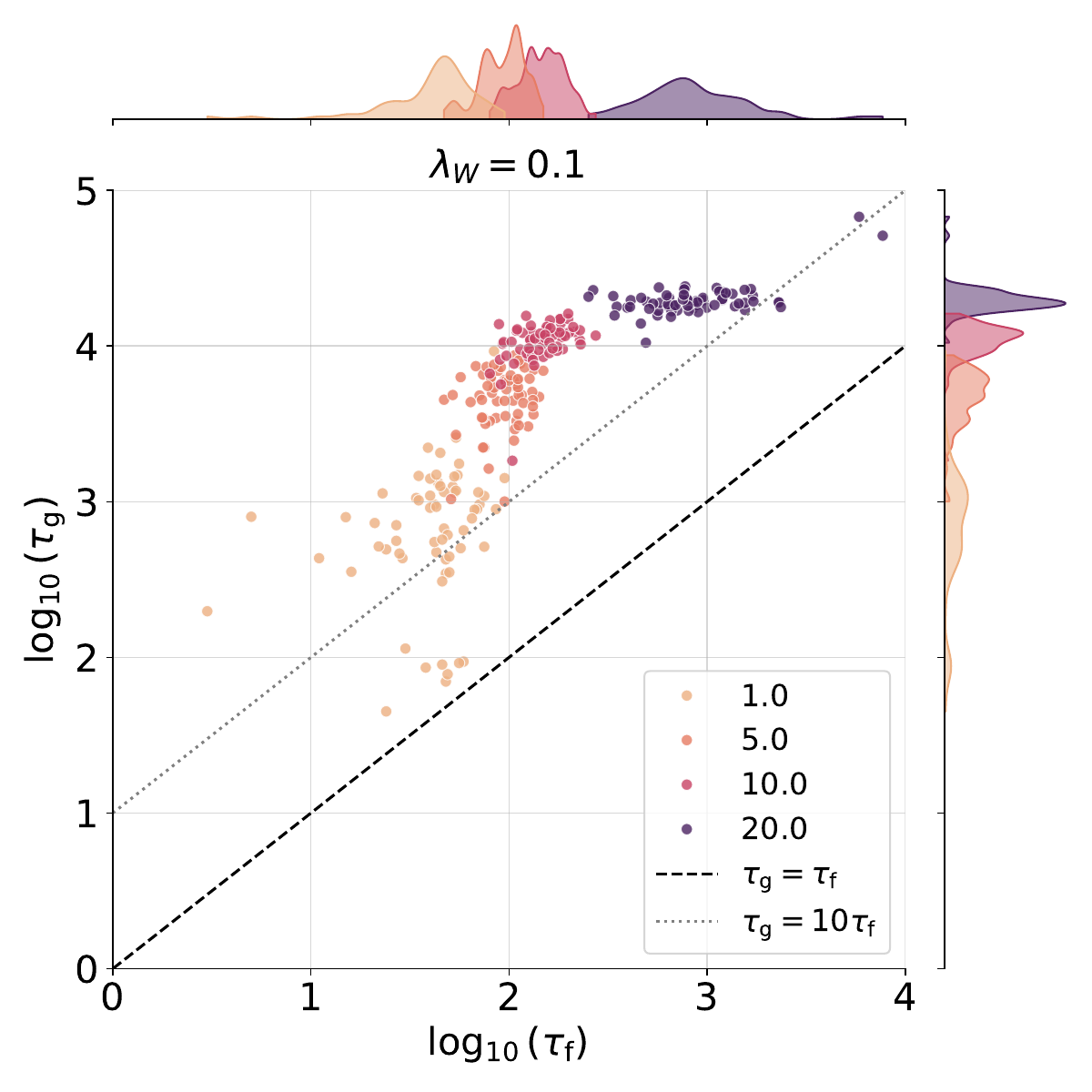}
        \caption{}
        \label{fig:temporal_dynamics_wd_1}
    \end{subfigure}%
    \begin{subfigure}{0.33\textwidth}
        \includegraphics[width=\linewidth]{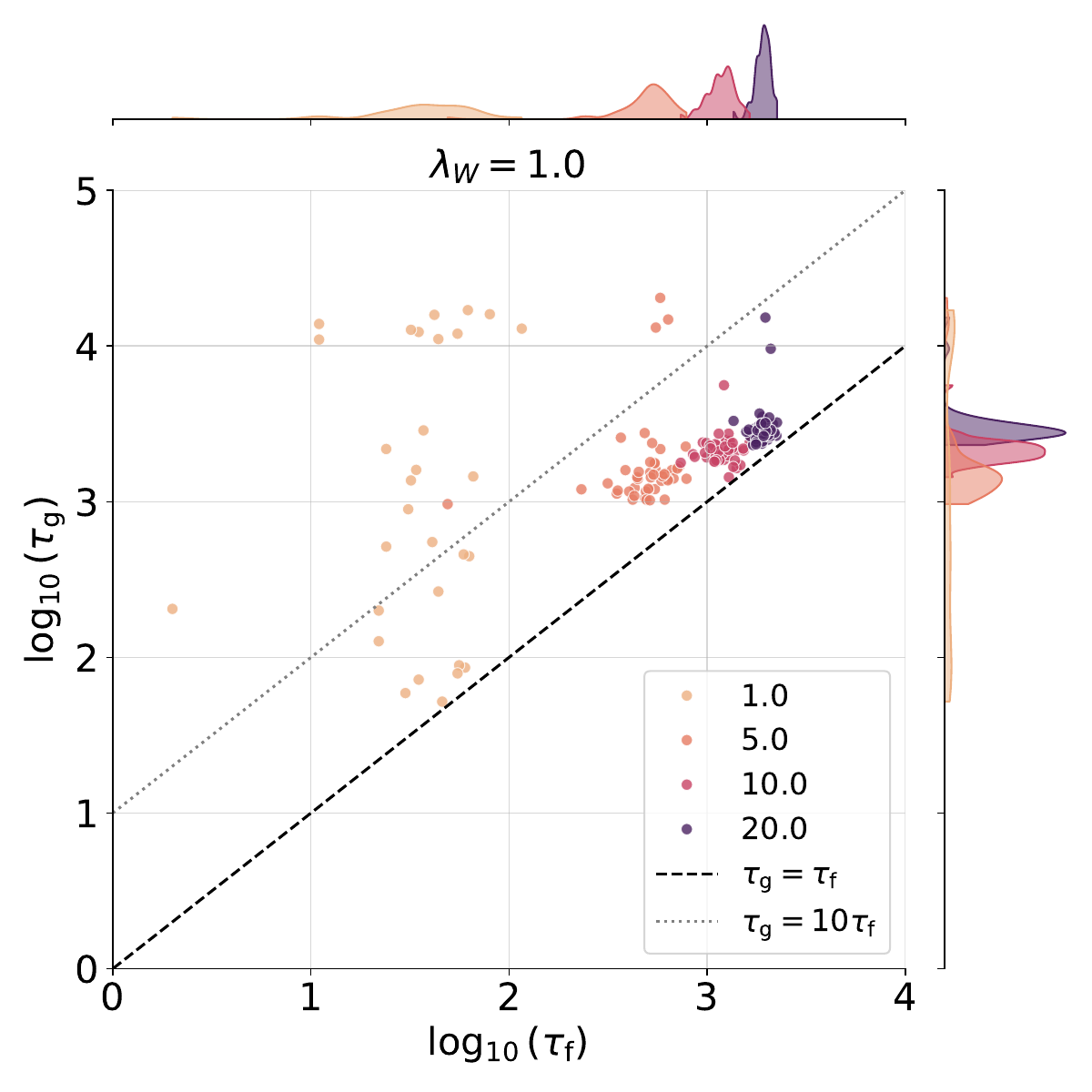}
        \caption{}
        \label{fig:temporal_dynamics_wd_2}
    \end{subfigure}
    \caption{Temporal dynamics of the memorization-to-generalization transition. (a) Violin plot tracking the absolute time gap $\log_{10}(\tau_\mathrm{g} - \tau_\mathrm{f})$ across different learning rates $\log_{10}(\eta) \in \{-5, \dots, 0\}$. Joint scatter and marginal density plots mapping generalization time $\log_{10}(\tau_\mathrm{g})$ against fitting time $\log_{10}(\tau_\mathrm{f})$ under varying weight initialization scales $\sigma_{\bm\theta} \in \{1.0, 5.0, 10.0, 20.0\}$ for (b) weak weight decay $\lambda_W = 0.1$ and (c) strong weight decay $\lambda_W = 1.0$. The dashed black line marks synchronous learning ($\tau_\mathrm{g} = \tau_\mathrm{f}$), while the dotted grey line marks the boundary threshold separating comprehension from grokking ($\tau_\mathrm{g} = 10\tau_\mathrm{f}$).}
\end{figure}

\noindent To fully characterize the transition from memorization to generalization, we evaluate the precise epoch at which these phases occur. We define the fit time $\tau_\mathrm{f}$ as the epoch where the training error drops below 0.1, and the generalization time $\tau_\mathrm{g}$ as the epoch where the test error drops below 0.1~\cite{grokking_representation_learning}. \\

\noindent Our results indicate that the explicit regularization strength $\lambda$ does not significantly alter the temporal distribution of the relative time gap $\tau_\mathrm{g} / \tau_\mathrm{f}$. Its primary effect is optimizing the overall success rate of runs that manage to generalize successfully. The foundational driving mechanism that induces the grokking transition in this overparameterized regime remains the implicit regularization generated by the AdamW optimizer's decoupled weight decay. \\

\noindent As illustrated in~\autoref{fig:temporal_dynamics_lr}, adjusting the learning rate $\eta$ dramatically shifts the absolute time gap $\tau_\mathrm{g} - \tau_\mathrm{f}$. Increasing the learning rate from $\log_{10}(\eta) = -5$ to $-1$ monotonically accelerates the onset of generalization, dragging the median delay down by nearly three orders of magnitude. However, once the learning rate reaches the extreme threshold of $\log_{10}(\eta) = 0$, this trend sharply inverts; excessive gradient steps introduce severe stochastic instability into the optimization trajectory, widening the variance and shifting the median delay back upward. \\

\noindent Furthermore, we investigate the joint impact of the weight initialization scale $\sigma_{\bm\theta}$ and weight decay $\lambda_W$ in~\autoref{fig:temporal_dynamics_wd_1} and~\autoref{fig:temporal_dynamics_wd_2}. When operating with a weaker weight decay ($\lambda_W = 0.1$), models initialized with a large parameter spread ($\sigma_{\bm\theta} \ge 10.0$, indicated by purple and pink hues) cluster highly in the upper-right corner of the scatter plot. These configurations require prolonged exploration of the flat loss landscape, fitting the training set late ($\log_{10}(\tau_\mathrm{f}) > 2$) and delaying the grokking transition close to the maximum training horizon. Crucially, as shown in~\autoref{fig:temporal_dynamics_wd_2}, increasing the weight decay strength to $\lambda_W = 1.0$ dramatically compresses these initialization trajectories. Under stronger implicit regularization, the high-$\sigma_{\bm\theta}$ clusters migrate downward, noticeably accelerating $\tau_\mathrm{g}$. However, this acceleration comes at a distinct cost to total convergence reliability. A close inspection of the marginal distributions reveals significantly lower counts for the $\sigma_{\bm\theta} = 20.0$ and $\sigma_{\bm\theta} = 10.0$ runs under $\lambda_W = 1.0$ compared to $\lambda_W = 0.1$. This indicates that while strong weight decay accelerates generalization for favorable seeds, it simultaneously suppresses the global success rate for extreme initializations by over-constraining the parameters, forcing a notable portion of runs into non-generalizing states before they can resolve the decision boundary. This trade-off offers a nuanced refinement to the traditional QML wisdom of utilizing narrow initializations ($\sigma_{\bm\theta} \to 0$) or specialized warm-starts to circumvent Barren Plateaus. While a high-capacity, broadly initialized QNN can navigate toward generalizing solutions when guided by implicit weight decay, the regularization strength must be carefully tuned: a moderate value promotes accelerated grokking, whereas an excessively aggressive decay over-corrects and chokes out generalization entirely.

\section{Discussion}

\noindent In this work, we provide the first empirical observation of both the grokking transition and epoch-wise double descent in QNNs. We demonstrate these phenomena on 2-qubit QNNs parameterized over the complete \textit{SU}(4) manifold. Our findings demonstrate that the optimization process within the flat regimes typical of overparameterized quantum circuits is far from static. Even during extended training phases where the training error is securely at zero and the test error remains completely stagnant, the model’s internal state is highly active. It continuously shifts between regions of differing algorithmic stability, weight-norm magnitudes, and internal representation structures, steadily evolving until the sudden onset of generalization occurs. \\

\noindent Crucially, we connected the grokking transition directly to algorithmic stability theory, demonstrating that the sudden drop in test error mirrors a sharp transition in the leave-one-out hypothesis stability ($\beta_{loo}$) and a compression of the model's internal representation space. While classical deep learning literature frequently frames grokking as an implicit competition between sparse and dense architectural sub-networks, our microscopic analysis reveals a distinct empirical manifestation in the frequency domain. The grokking transition is characterized by a structural phase-realignment of the learned model function, where diffuse, chaotic spectral coefficients collapse into highly symmetric, sparse, and phase-aligned Fourier harmonics. However, this high-capacity regime comes with a major vulnerability, where the model's generalization performance decays in the post-grokking regime. Without structural anchors, the unconstrained drift of the weight-norm in late-stage training pulls the optimizer out of generalizing solutions and into highly complex, overfitted regions of the Hilbert space. \\

\noindent Our investigation into the temporal characterization of grokking demonstrates that QNNs exhibit higher generalization success when operated in the heavily overparameterized regime. This delayed memorization-to-generalization transition is driven heavily by implicit regularization. Paradoxically, despite a large initialization scale ($\sigma_{\bm\theta} \ge 10.0$), these models can achieve rapid generalization times if guided by a suitable implicit weight decay. This finding stands in stark contrast to prevailing paradigms in the QML community, which traditionally advocate for narrow initialization scales ($\sigma_{\bm\theta} \to 0$) or specialized warm-starts to circumvent Barren Plateaus. We show that narrow initializations are not a universal requirement for generalization. Instead, a broad parameter exploration space can successfully navigate toward generalizing solutions, provided its trajectory is regularized by appropriate weight-decay constraints. \\

\noindent To combat late-stage generalization decay, we established that incorporating a weak explicit regularizer (such as the Lipschitz bound or an $L_2$ norm constraint) to the loss function is essential. This acts as a structural anchor, stabilizing the grokking phase and ensuring the QNN retains its generalization capabilities across indefinitely long training horizons without altering the underlying temporal onset of the transition\\

\noindent Several compelling avenues emerge from this study. First, scaling these architectures to larger multi-qubit systems will allow us to assess whether the grokking timeline ($\tau_g / \tau_f$) compresses or expands with larger state-spaces. Second, evaluating these phenomena against noisy gradients or simulated hardware noise will determine if physical decoherence acts as a natural regularizer or disrupts the stability of the grokking phase entirely.

\section{Acknowledgment}

This work was funded by the Dieter Schwarz Stiftung within the Heilbronn Forschungs und Innovationszentrum (HNFIZ). D.P thanks Marie Kempkes and Luca Oneto for helpful discussions.  

\bibliographystyle{unsrt}
\bibliography{sample}

\begin{appendices}

\section{Tunable problem difficulty with tighter boundary samples}
\label{app:groksistency_d2b}

\begin{figure}[h]
    \centering
    \begin{subfigure}{0.5\textwidth}
        \includegraphics[width=\linewidth]{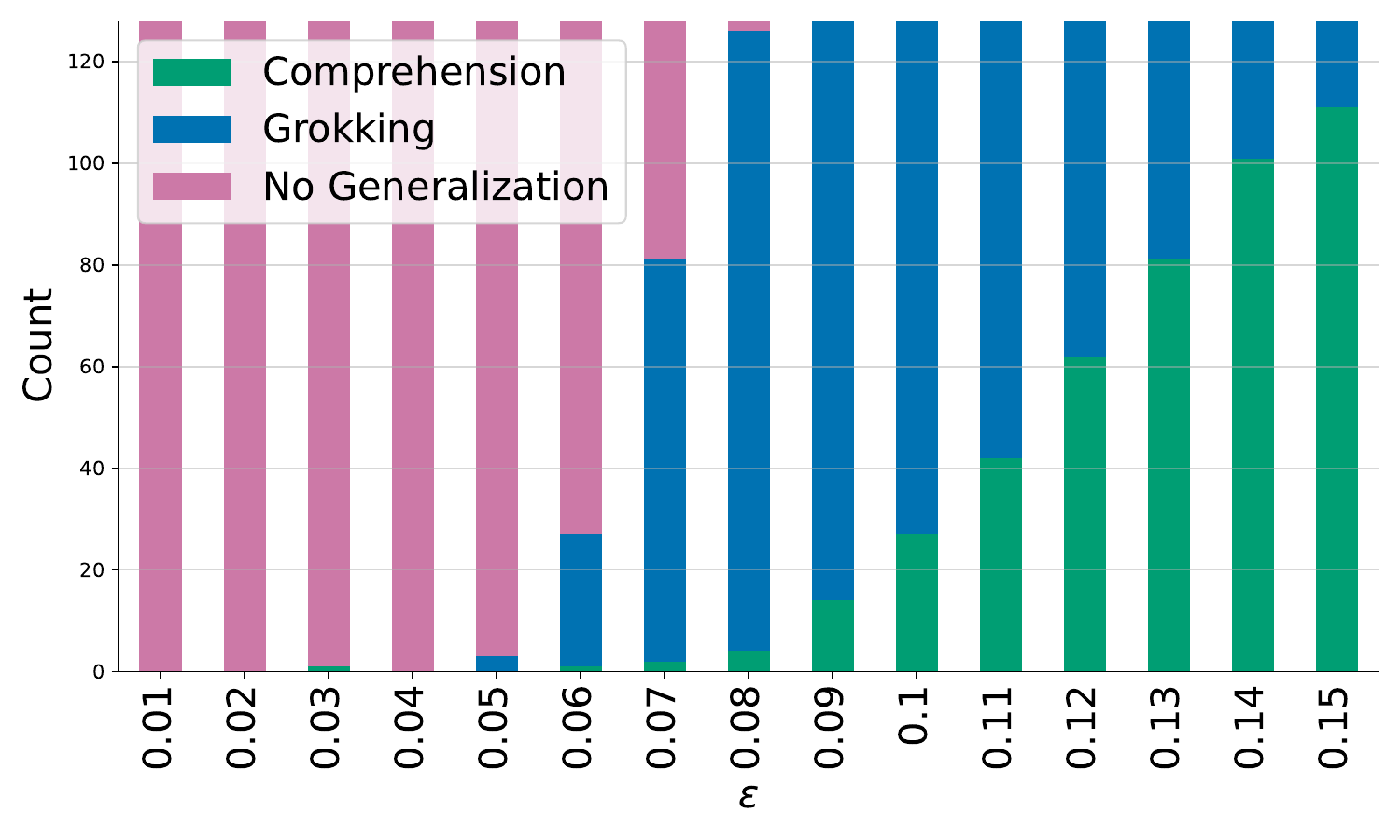}
        \caption{}
        \label{fig:groksistency_d2b_phase_ratios}
    \end{subfigure}%
    \begin{subfigure}{0.5\textwidth}
        \includegraphics[width=\linewidth]{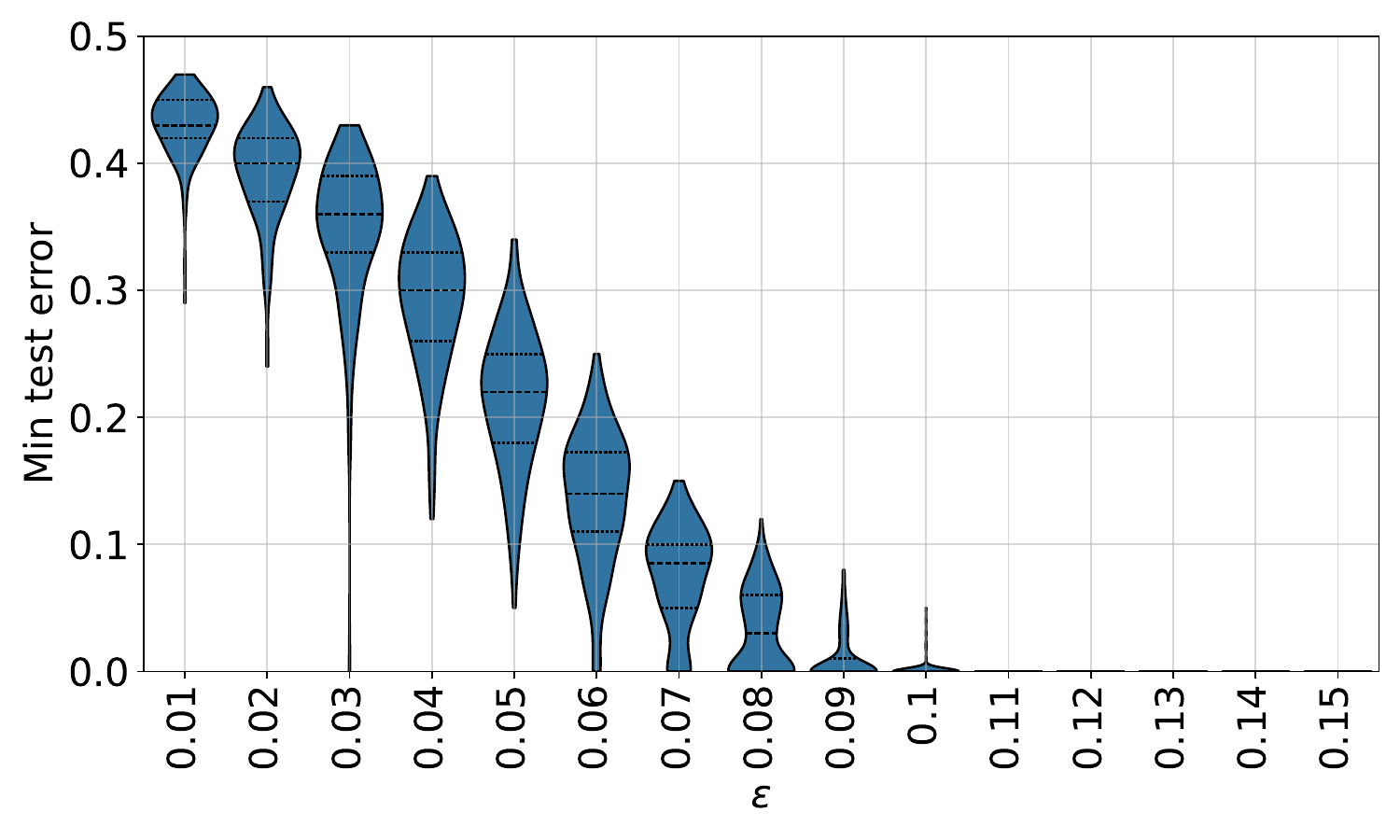}
        \caption{}
        \label{fig:groksistency_d2b_final_test_error}
    \end{subfigure}
    \begin{subfigure}{0.5\textwidth}
        \includegraphics[width=\linewidth]{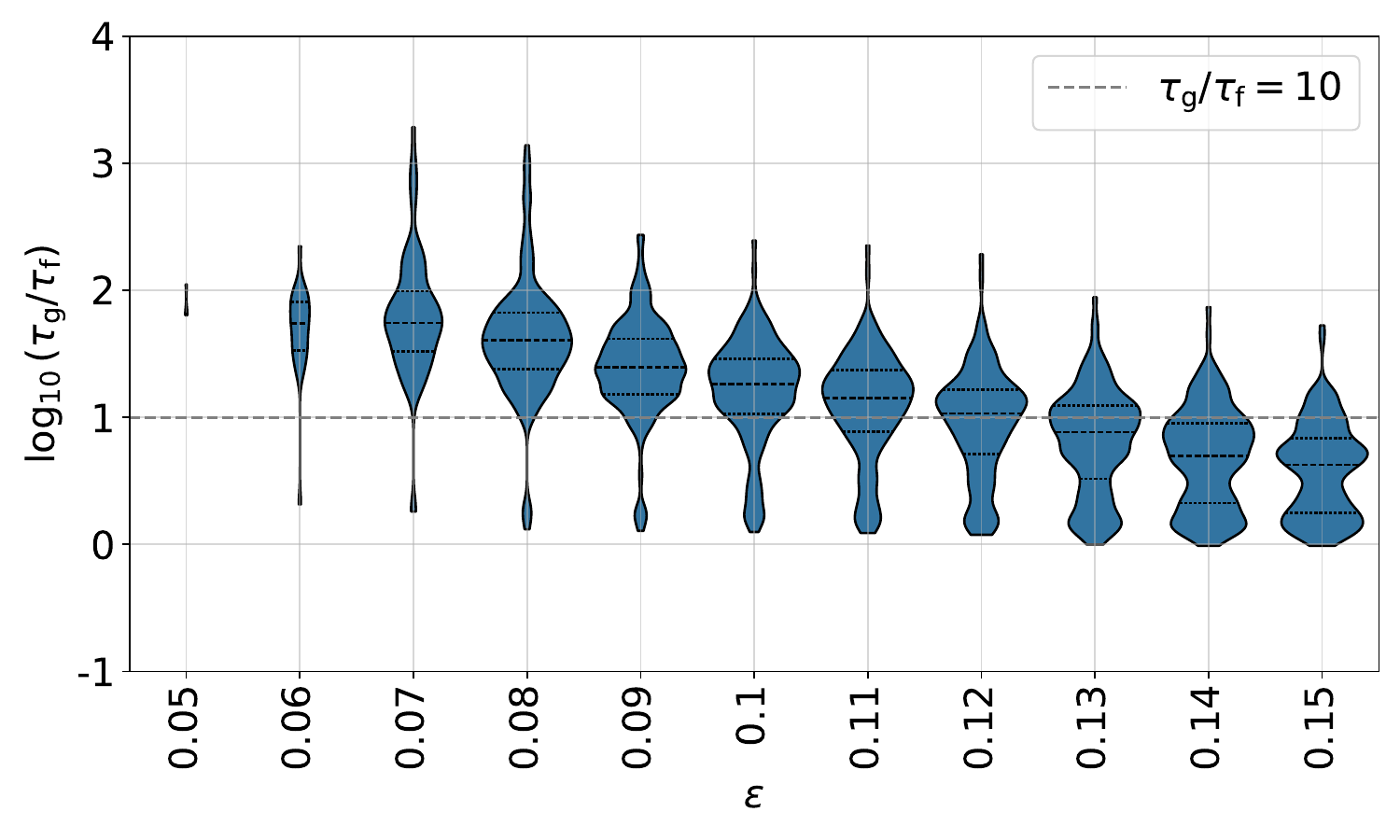}
        \caption{}
        \label{fig:groksistency_d2b_tratio}
    \end{subfigure}%
    \begin{subfigure}{0.5\textwidth}
        \includegraphics[width=\linewidth]{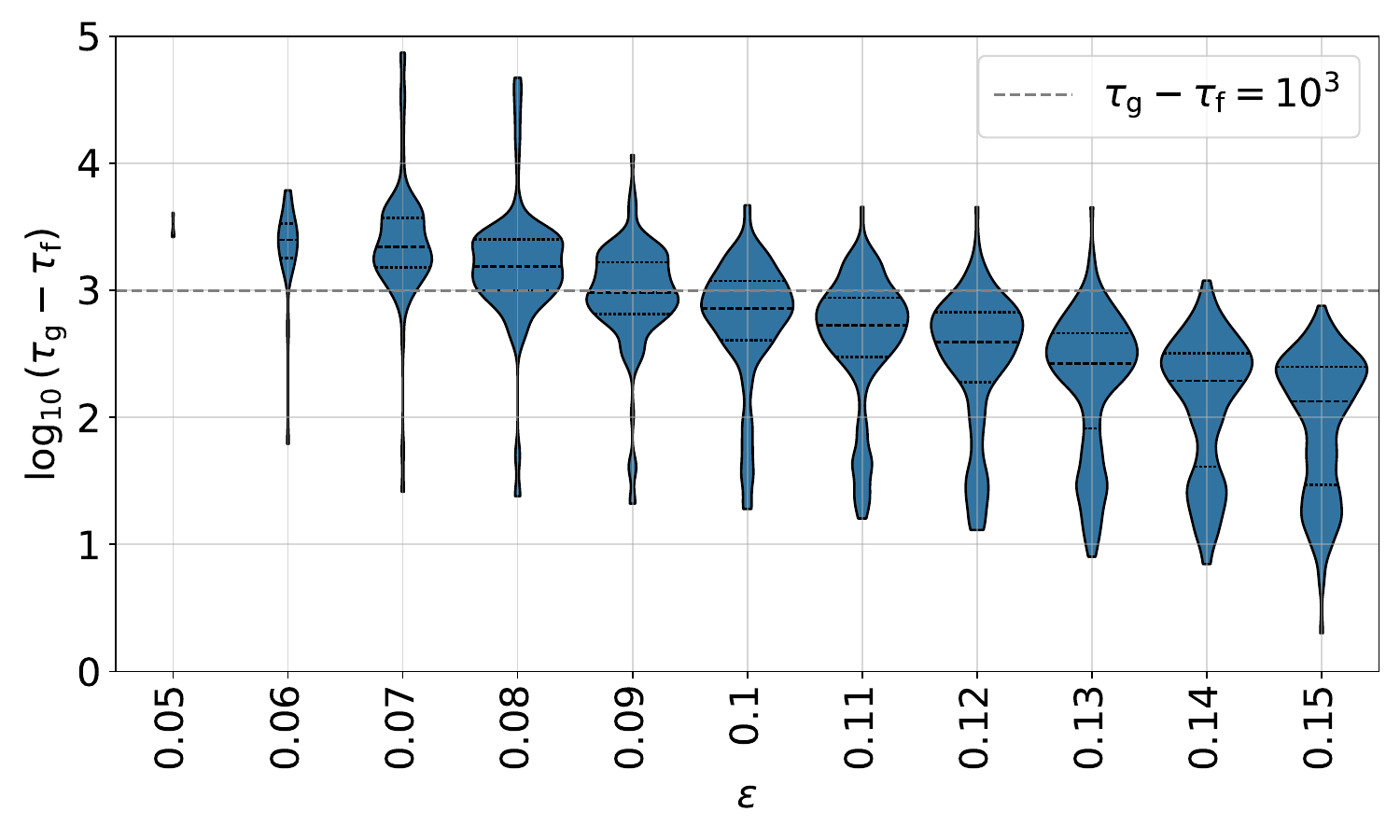}
        \caption{}
        \label{fig:groksistency_d2b_tdelta}
    \end{subfigure}
    \caption{Behavior of the quantum learner evaluated over 128 independent training runs for each boundary distance parameter $\varepsilon \in \{0.01, \dots 0.15\}$ and $10^5$ epochs.  (a) Barplot detailing the distribution of learning phases. (b) Violin plot illustrating the distribution of the minimum test error from the whole training history. (c) Relative time gap ratio ($\tau_\mathrm{g} / \tau_\mathrm{f}$) and (d) absolute temporal delay ($\tau_\mathrm{g} - \tau_\mathrm{f}$) until the grokking transition, conditioned strictly on the subset of successfully generalizing runs. Note that the horizontal lines in panels (c) and (d) signify the threshold criteria distinguishing phases according to~\autoref{tab:phase_categorization}.}
    \label{fig:groksistency_d2b}
\end{figure}

\noindent In our formulation, the structural difficulty of the learning problem can be explicitly controlled by modulating $\varepsilon > 0$, which defines the margin or minimum distance of the test samples to the SVM decision boundary. Decreasing $\varepsilon$ places test evaluation points closer to the classification interface, drastically reducing the geometric tolerance for suboptimal decision boundaries and increasing the required precision of the learned representations. Consequently, we hypothesize that tighter boundary margins will increase the generalization delay. This will manifest as a widening of both the relative time gap ratio $\tau_\mathrm{g} / \tau_\mathrm{f}$ and the absolute difference $\tau_\mathrm{g} - \tau_\mathrm{f}$, while simultaneously degrading the final test performance. \\

\noindent Our empirical findings across 128 independent random seeds for each increment of $\varepsilon$ are illustrated in~\autoref{fig:groksistency_d2b} and strongly validate this hypothesis. As shown in the phase distribution barplot (\autoref{fig:groksistency_d2b_phase_ratios}), varying $\varepsilon$ induces a systematic, monotonic transition between macro-learning behaviors. In the large-margin regime ($\varepsilon = 0.15$), a substantial fraction of the initializations converge via clean \textit{Comprehension} (green), meaning the boundary is loose enough for early-epoch representations to generalize immediately. However, as $\varepsilon$ decreases, the proportion of comprehension runs steadily shrinks, disappearing entirely when $\varepsilon \le 0.05$. This compression of the comprehension phase is accompanied by a massive expansion of the \textit{Grokking} domain (yellow). Because the target boundary requires immense harmonic alignment to resolve tight margins, the QNN is forced into an extended memorization plateau before finding the generalizing solution. At the extreme difficulty limit ($\varepsilon = 0.01$), the optimization landscape becomes so constrained that a notable portion of seeds fail to generalize within the training horizon, registering as \textit{No Generalization} (red). \\

\noindent This geometric sensitivity is further illuminated by the minimum test error distributions in~\autoref{fig:groksistency_d2b_final_test_error}. For relatively relaxed margins ($\varepsilon \ge 0.1$), the violin plots are highly concentrated with their mass localized tightly at zero error, indicating robust generalization across almost all seeds. \\
As $\varepsilon$ drops below $0.1$, the distributions visibly widen and develop upward tails toward higher test errors. This structural broadening indicates that even when a grokking transition is triggered, the precision of the emerging phase-aligned representations fluctuates under stochastically driven initial conditions. When the test points are clustered tightly against the boundary, even minor deviations in the QNN's learned hyperplane result in severe test classification penalties, cementing $\varepsilon$ as a highly predictable, tunable proxy for task difficulty. \\

\noindent To directly isolate the temporal scales of this phenomenon,~\autoref{fig:groksistency_d2b_tratio} and~\autoref{fig:groksistency_d2b_tdelta} map the relative $\tau_\mathrm{g} / \tau_\mathrm{f}$ and absolute time delay $\tau_\mathrm{g} - \tau_\mathrm{f}$ as functions of $\varepsilon$. In line with our core prediction, tightening the decision margin from $\varepsilon = 0.15$ down to $0.03$ forces an exponential spike in the alignment timescale. The median relative time ratio (\autoref{fig:groksistency_d2b_tratio}) sweeps upward by orders of magnitude, moving from near-synchronous comprehension timelines to ratios far exceeding $10^2$. This is mirrored in the absolute delay profile (\autoref{fig:groksistency_d2b_tdelta}), where the span of the static memorization plateau stretches extensively toward $10^4$ epochs. Interestingly, at the most rigid difficulty tier ($\varepsilon = 0.01$), the upward trend exhibits a visual inversion, manifesting as a slight contraction in both the median ratio and absolute delay variance. Rather than indicating that the optimization problem eases, this structural change is a direct artifact of survivor bias within a finite training horizon. As verified by the phase counts in~\autoref{fig:groksistency_d2b_phase_ratios}, $\varepsilon = 0.01$ introduces a notable portion of runs that completely fail to generalize before the maximum training epoch is reached. Because these chronically delayed runs are categorized under \textit{No Generalization}, they are naturally omitted from the conditional tracking in~\autoref{fig:groksistency_d2b_tratio} and~\autoref{fig:groksistency_d2b_tdelta}. The remaining successful pool is consequently skewed toward stochastically accelerated initialization seeds, bounding the apparent distribution.

\section{Computation of Fourier coefficients of the quantum learner from~\autoref{eq:model}}
\label{app:fourier_computation}

\noindent The Fourier-coefficients $c_{n_1, n_2}(\bm\theta_t)$ of the model $f(\bm{x}; \bm\theta_t)$ at time $t$, see~\autoref{eq:model} are obtained by

\begin{align}
    c_{n_1, n_2}(\bm\theta_t) &= \frac{1}{L^2} \iint \, f(x_1, x_2;\bm\theta_t) e^{-i \frac{2\pi}{L} \left( x_1 n_1 + x_2 n_2 \right)} \, \mathrm{d}x_1 \mathrm{d}x_2 \, , \label{eq:definition_cnn} \\
    &\approx \frac{1}{N} \sum\limits_{j=1}^N \, f\left(x_1^{(j)}, x_2^{(j)}; \bm\theta_t\right) \, e^{-i \frac{2 \pi}{L}\left( x_1 n_1 + x_2 n_2 \right)} \, ,
    \label{eq:cnn}
\end{align}
where $L = \max\{ L_1, L_2 \}, L_i = \max\limits_j\left\{ x_i^{(j)} \right\} - \min\limits_j\left\{x_i^{(j)}\right\}$ is the length of the square that contains all training samples $x^{(j)}$ and $\mathbf{x} = (x_1, x_2)^T$ denotes the spatial coordinate vector of the grid sample. For our results in~\autoref{fig:hero_fourier}, $f(\bm{x}; \bm\theta_t)$ was sampled on a $N_1 \times N_2$ grid, where $N_1 = N_2 =256, N = 256^2$ for $t \in \{ 0, \tau_\mathrm{f}, \tau^{\pm}_\mathrm{g}, 10^5 \}$. \\

\noindent The dramatic structural re-alignment observed during the grokking transition can be rigorously understood by examining the \textit{idealized} target function of the concentric circle dataset. Given that the optimal decision boundary is a circle of radius $R$ separating labels $y \in \{-1, +1\}$ inside a square domain of side length $L$ centered at the origin, the ideal classifier maps to a radially symmetric step function
\begin{equation}
f(\mathbf{x}) = 2 \cdot \mathbb{I}(\lVert\mathbf{x}\rVert \le R) - 1,
\label{eq:ideal_classifier}
\end{equation}
where $\mathbb{I}$ is the indicator function. By utilizing a uniform grid discretization of $N= N_1 N_2$ total samples over the domain $[0, L]^2$, the discrete Fourier coefficients $c_{n_1, n_2}$ evaluated at integer frequencies $\mathbf{n} = (n_1, n_2)^T$ can be split via linearity into a foreground disk component and a background offset component
\begin{equation}
\label{eq:discrete_split}
    c_{n_1, n_2} = \frac{2}{N} \sum_{\mathbf{x}^{(j)} \in \mathcal{D}_R} \exp\left(-i \frac{2\pi}{L} \mathbf{n} \cdot \mathbf{x}^{(j)}\right) - \frac{1}{N} \sum_{j=1}^{N} \exp\left(-i \frac{2\pi}{L} \mathbf{n} \cdot \mathbf{x}^{(j)}\right),
\end{equation}
where $\mathcal{D}_R = \{ \mathbf{x}^{(j)} \mid \lVert\mathbf{x}^{(j)}\rVert \le R \}$ represents the subset of grid points falling strictly inside the circular boundary. The second term represents a uniform exponential sum over a complete, orthogonal discrete lattice. For any integer frequencies $n_1, n_2$, this sum factorizes into two independent one-dimensional geometric series
\begin{equation}
\frac{1}{N} \sum_{j=1}^{N} \exp\left(-i \frac{2\pi}{L} \mathbf{n} \cdot \mathbf{x}^{(j)}\right) = \left( \frac{1}{N_1} \sum_{j_1=1}^{N_1} e^{-i \frac{2\pi}{L} n^{ }_1 x_1^{(j_1)}} \right) \left( \frac{1}{N_2} \sum_{j_2=1}^{N_2} e^{-i \frac{2\pi}{L} n^{ }_2 x_2^{(j_2)}} \right).
\end{equation}
Due to the orthogonality of the discrete Fourier basis over a full period, this product evaluates exactly to a Kronecker delta function
\begin{equation}
\frac{1}{N} \sum_{j=1}^{N} \exp\left(-i \frac{2\pi}{L} \mathbf{n} \cdot \mathbf{x}^{(j)}\right) = \delta_{n_1, 0} \, \delta_{n_2, 0} = \begin{cases} 1 & \text{if } n_1 = 0 \text{ and } n_2 = 0, \\ 0 & \text{otherwise}. \end{cases}
\end{equation}

\noindent The first term sums exclusively over the grid points residing inside the radius $R$. 
\begin{enumerate}
    \item \textbf{For the Zero-Frequency Component ($n_1 = 0, n_2 = 0$):}
    The complex exponential collapses to unity ($e^0 = 1$). The sum simply counts the total number of discrete points falling inside the disk, $N_{\mathcal{D}_R} = |\mathcal{D}_R|$
    \begin{equation}
    \frac{2}{N} \sum_{\mathbf{x}^{(j)} \in \mathcal{D}_R} (1) = 2 \frac{N_{\mathcal{D}_R}}{N} \approx 2 \left( \frac{\pi R^2}{L^2} \right).
    \end{equation}
    Here, $\frac{\pi R^2}{L^2}$ represents the geometric ratio of the circle's area to the total square domain area.
    \item \textbf{For Non-Zero Frequencies ($\lVert\mathbf{n}\rVert \neq 0$):}
    When the grid density is sufficiently high (large $N$), the Riemann sum over the bounded disk domain $\mathcal{D}_R$ cleanly converges to its continuous integral equivalent. Recalling that the area element of a single grid cell is $\Delta A = \frac{L^2}{N}$, we substitute $\frac{1}{N} = \frac{\Delta A}{L^2}$
    \begin{equation}
    \frac{2}{N} \sum_{\mathbf{x}^{(j)} \in \mathcal{D}_R} e^{-i \frac{2\pi}{L} \mathbf{n} \cdot \mathbf{x}^{(j)}} \approx \frac{2}{L^2} \iint_{\lVert\mathbf{x}\rVert \le R} e^{-i \frac{2\pi}{L} \mathbf{n} \cdot \mathbf{x}} \, \mathrm{d}x_1 \mathrm{d}x_2. \label{eq:sum2riemann_cnn}
    \end{equation}
    We exploit the radial symmetry by transforming to polar coordinates. Let
    \begin{align}
    x_1 &= r \cos\phi, \quad &x_2 &= r \sin\phi, \quad &\mathrm{d}x_1 \mathrm{d}x_2 &= r \, \mathrm{d}r \mathrm{d}\phi, \\
    n_1 &= n \cos\vartheta, \quad &n_2 &= n \sin\vartheta, \quad &n &= \lVert\boldsymbol{n}\rVert = \sqrt{n_1^2 + n_2^2}.
    \end{align}
    The inner vector product in the complex exponential becomes
    \begin{equation}
    \boldsymbol{n} \cdot \mathbf{x} = n_1 x_1 + n_2 x_2 = n r (\cos\vartheta \cos\phi + \sin\vartheta \sin\phi) = n r \cos(\phi - \vartheta).
    \end{equation}
    Due to the periodic nature of the cosine function over a full $2\pi$ cycle, we can shift the integration variable to $\alpha = \phi - \vartheta$. The integral over the angle matches the standard integral definition of the Bessel function of the first kind of order zero, $J_0(z) = \frac{1}{2\pi} \int_{0}^{2\pi} e^{-i z \cos\alpha} \mathrm{d}\alpha$
    \begin{equation}
    \int_{0}^{2\pi} e^{-i \frac{2\pi}{L} n r \cos\alpha} \mathrm{d}\alpha = 2\pi J_0\left(\frac{2\pi n r}{L}\right).
    \end{equation}
    Substituting this back reduces the two-dimensional Fourier integral into a one-dimensional Hankel transform of order zero
    \begin{equation}
    \hat{g}(\boldsymbol{n}) = 2\pi \int_{0}^{R} r J_0\left(\frac{2\pi n r}{L}\right) \, \mathrm{d}r.
    \end{equation}
    To evaluate this remaining integral, we utilize the standard mathematical identity for Bessel functions, $\frac{\mathrm{d}}{\mathrm{d}z} [z J_1(z)] = z J_0(z)$, which yields the integration rule $\int z J_0(z) \mathrm{d}z = z J_1(z)$. Let us perform a change of variables by setting $z = 2\pi n r/L$, which implies $\mathrm{d}r = L \mathrm{d}z / 2\pi n$. The integration limits shift from $[0, R]$ to $[0, 2\pi n R/L]$
    \begin{equation}
    \hat{g}(\boldsymbol{n}) = 2\pi \int_{0}^{\frac{2\pi n R}{L}} \left(\frac{L z}{2\pi n}\right) J_0(z) \frac{L \, \mathrm{d}z}{2\pi n} = \frac{L^2}{2\pi n^2} \int_{0}^{\frac{2\pi n R}{L}} z J_0(z) \, \mathrm{d}z.
    \end{equation}
    Applying the integration identity gives
    \begin{equation}
    \hat{g}(\boldsymbol{n}) = \frac{L^2}{2\pi n^2} \Big[ z J_1(z) \Big]_{0}^{\frac{2\pi n R}{L}} = \frac{L^2}{2\pi n^2} \Big( \frac{2\pi n R}{L} J_1\left( \frac{2\pi n R}{L} \right) - 0 \Big),
    \end{equation}
    where we have used the fact that $\lim_{z \to 0} z J_1(z) = 0$. Simplifying terms isolates the final expression for the disk's contribution
    \begin{equation}
    \hat{g}(\boldsymbol{n}) = \frac{L R J_1\left( \frac{2\pi R n}{L}\right)}{n}.
    \end{equation}
    Multiplying by $2/L^2$ (from~\autoref{eq:sum2riemann_cnn}) yields the result for $\lvert\lvert \bm{n} \rvert\rvert \neq 0$
    \begin{equation}
        c_{n_1,n_2} = \frac{2R}{L} \frac{J_1\left(2 \pi \frac{R}{L} n \right)}{n} \, .
    \end{equation}
\end{enumerate}
Combining the zero-frequency offset and the non-zero frequency harmonic dampening yields the full analytical description for the discrete heatmap coordinates
\begin{equation}
c_{n_1, n_2} \approx \begin{cases} 
2 \left( \dfrac{\pi R^2}{L^2} \right) - 1 & \text{if } n_1 = 0 \text{ and } n_2 = 0, \\[12pt]
\dfrac{2R}{L} \dfrac{J_1\left(2\pi \dfrac{R}{L} \sqrt{n_1^2 + n_2^2}\right)}{\sqrt{n_1^2 + n_2^2}} & \text{otherwise}.
\end{cases}
\end{equation}
The algebraic expression $J_1(2\pi R \lVert\boldsymbol{n}\rVert/L)/\lVert\boldsymbol{n}\rVert$ describes a classic Airy-like diffraction pattern characterized by a prominent central maximum surrounded by concentric rings of oscillating intensity. Because the entire radial envelope of these coefficients decays asymptotically as $\mathcal{O}(\lVert\mathbf{n}\rVert^{-3/2})$, this baseline perfectly dictates the emergence of the sharp rings and the fading outward amplitude observed empirically in the post-grokking magnitude spectrum. \\

\noindent Crucially, this analytical baseline provides a complete physical explanation for the evolution of the complex phase profile ($\arg(c_{n_1, n_2})$) across different training regimes. Because the idealized step function is purely real-valued and perfectly symmetric about the origin, its theoretical Fourier coefficients must be strictly real. In the complex plane, a purely real number is constrained to a binary phase profile: it possesses a phase of exactly $0$ where the value is positive, and flips sharply to $\pm\pi$ where the value is negative. 
Consequently, the phase landscape is entirely dictated by the alternating algebraic sign of the oscillating Bessel function $J_1(z)$ as its radial argument increases. This mathematical constraint beautifully illuminates the three distinct phases of the QNN training dynamics:
\begin{itemize}
    \item \textbf{Initial to Pre-Grokking Phase:} During early training, the model relies on unaligned, high-frequency parameter configurations to memorize individual training coordinates. This jagged, asymmetric landscape generates fully complex Fourier coefficients with arbitrary real and imaginary parts, producing the highly chaotic and scattered phase distributions observed initially.
    \item \textbf{Towards the Post-Grokking Phase:} Upon entering the generalization regime, the combined regularizing effects of optimization and weight decay force the quantum circuit to strip away overfitted high-frequency modes. The QNN restricts its operation to a low-frequency harmonic manifold that mirrors the target function. As the network's output converges toward the real-valued circular baseline, the coefficients are driven onto the real axis of the complex plane. The phase spectrum cleanly reorganizes into highly structured, concentric bands of alternating $0$ and $\pi$ phases, serving as a direct visual signature of the quantum state aligning with the roots of the classical Bessel function.
    \item \textbf{Late-Stage Generalization Decay:} In the final stages of unconstrained training, as the parameter norms swell, the model begins to introduce subtle structural asymmetries and superfluous phase variations into the state overlays to overfit residual noise profiles. Even a minute deviation from perfect radial symmetry pulls the Fourier coefficients off the real axis and back into the complex plane. This mathematical drift results in the visual blurring or ``smearing'' of the sharp phase boundaries, capturing the exact onset of overparameterization decay.
\end{itemize}

\section{Stability of the grokking phase}
\label{app:groksistency_lambda}

\begin{figure}[h]
    \centering
    \begin{subfigure}{0.5\textwidth}
        \includegraphics[width = \linewidth]{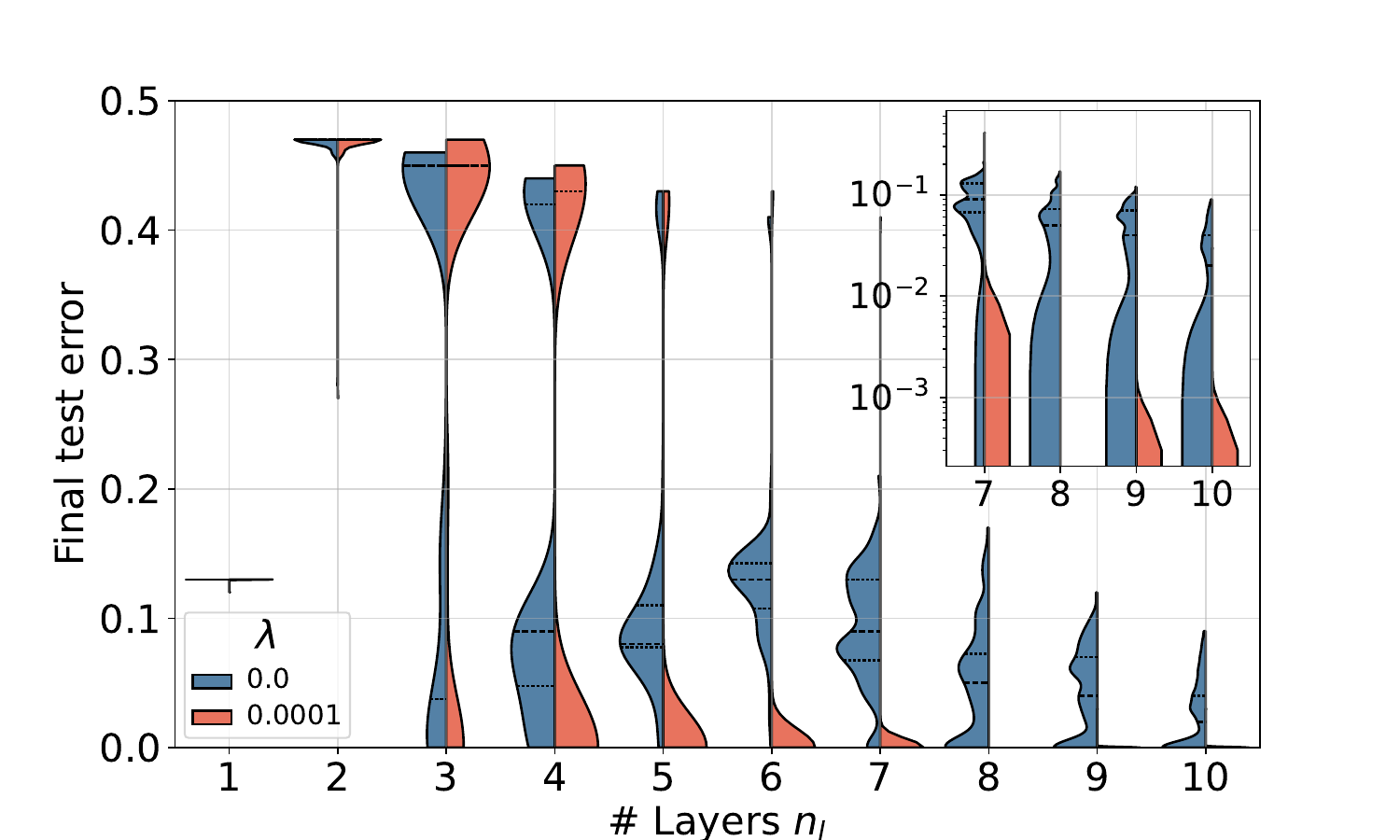}
        \caption{}
        \label{fig:groksistency_l2_final_test_error}
    \end{subfigure}%
    \begin{subfigure}{0.5\textwidth}
        \includegraphics[width= \linewidth]{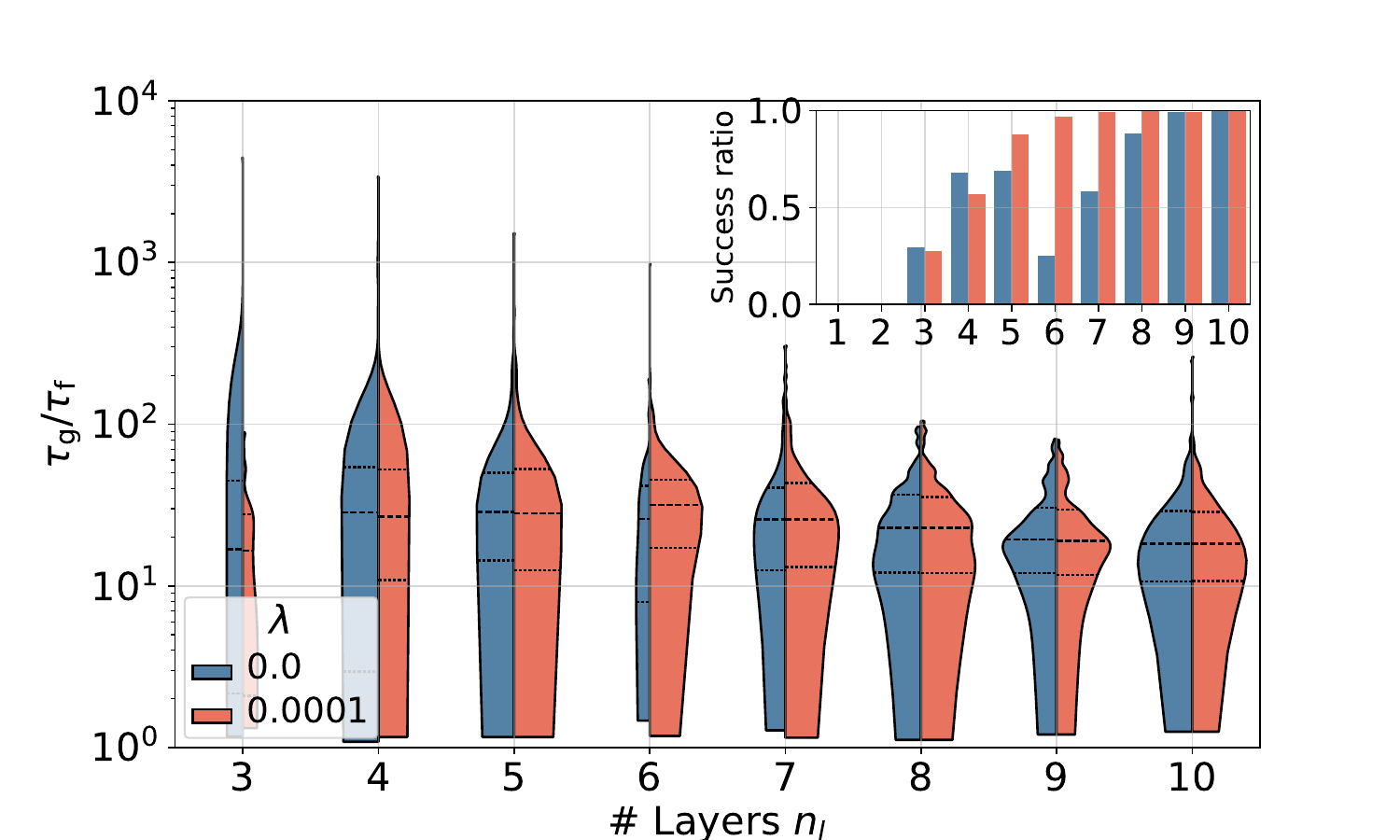}
        \caption{}
        \label{fig:groksistency_l2_delay_and_success}
    \end{subfigure}
    \caption{(a) Violin plot of the final test error across 128 independent training runs for each layer depth $n_l \in \{1, \dots, 10 \}$, comparing unregularized training ($\lambda=0$) against explicit weight-norm regularization ($\lambda=10^{-4}$) and $10^5$ epochs. The introduction of regularization drastically suppresses post-transition drift, leading to near-zero final test error for deeper architectures ($n_l \ge 4$). (b) The temporal gap ratio $\tau_\mathrm{g}/\tau_\mathrm{f}$ calculated exclusively for the subset of successful runs achieving a final test error below $0.1$. The inset depicts the absolute generalization success rate for each $n_l$ with and without explicit regularization.}
    \label{fig:groksistency_l2}
\end{figure}

\noindent Explicit weight-norm regularization $\lambda$ plays a vital role in bounding the model's parameters, effectively preventing the representation from drifting back toward high-complexity, non-generalizing solutions after the grokking transition has occurred. To evaluate the statistical robustness of this effect, we compare the final test error distribution of 128 independent training runs across depths $n_l \in \{1, \dots, 10\}$ in both the unregularized ($\lambda = 0$) and regularized ($\lambda = 10^{-4}$) cases.\\

\noindent Our empirical results, summarized in the violin plots of~\autoref{fig:groksistency_l2}, demonstrate that explicit weight-norm regularization substantially reinforces the stability of the generalized state. This effect is strongly coupled with QNN depth. For shallow models ($n_l < 4$), the presence of $\lambda = 10^{-4}$ yields no perceptible divergence from the unregularized baselines, as these architectures lack the capacity to sustain the grokking phase under these initialization conditions. Conversely, for deeper QNNs ($n_l > 4$), the regularized runs exhibit an overwhelming convergence toward a final test error of approximately zero. In contrast, the unregularized trajectories display significant variance, with a prominent cluster of runs drifting back to high test error regimes by the end of training. \\

\noindent To determine whether this regularization alters the inner temporal mechanics of the transition,~\autoref{fig:groksistency_l2_delay_and_success} tracks the phase time gap ratio $\tau_\mathrm{g}/\tau_\mathrm{f}$ for all successfully generalizing runs. Crucially, the overlapping distributions between $\lambda = 0$ and $\lambda = 10^{-4}$ indicate that a small, explicit weight decay penalty does not alter the absolute speed or delayed nature of the grokking transition itself. Instead, its primary mechanism is the preservation of the generalizing state once found. As shown in the inset of~\autoref{fig:groksistency_l2_delay_and_success}, this preservation leads to a massive, systematic boost in the absolute generalization success rate for all networks with $n_l > 4$.

\end{appendices}

\end{document}